# Differentially Private Hierarchical Heavy Hitters*

Ari Biswas, *University of Warwick*, tcs@randomwalks.xyz
Graham Cormode, *University of Oxford*, graham.cormode@cs.ox.ac.uk
Yaron Kanza, *AT&T Research*, yk140j@att.com
Divesh Srivastava, *AT&T Research*, divesh@research.att.com
Zhengyi Zhou, *AT&T Research*, zz547k@att.com

**Abstract**

The task of finding *Hierarchical* Heavy Hitters (HHH) was introduced by Cormode et al. [VLDB 2003] as a generalisation of the heavy hitter problem. While finding HHH in data streams has been studied extensively, the question of releasing HHH when the underlying data is private remains unexplored. In this paper, we study differentially private HHH release in both the streaming and non-streaming setting. In the non-streaming setting, we show the surprising result that the *relative error* in estimating the residual count for any prefix is *independent* of the height of the hierarchy and the number of heavy hitters in the stream. Meanwhile, in the streaming setting, although the exact version of HHH has low global sensitivity (as counting queries are 1-sensitive), the approximation functions due to streaming have high global sensitivity, linear in the available space. Despite this obstacle, we show that the absolute error for estimating frequencies in the steaming setting is independent of the available space.

## 1 Introduction

The task of finding *Heavy Hitters* (HH), a.k.a. frequent items, in a dataset is one of the most well-studied problems in data science. The task has been studied under the streaming model of computation [12, 13], distributed computation [10], and even through the lens of secure computation [15]. In this work, we adopt the lens of *differential privacy* (DP) to study the *Hierarchical Heavy Hitters* (HHH) problem, introduced by Graham Cormode, Flip Korn, Shanmugavelayutham Muthukrishnan, and Divesh Srivastava. [13] as a generalisation of the heavy hitter problem. The problem of DP-HHH is motivated by the observation that data is often both *hierarchical* and *confidential*. Consider checking for evidence of discrimination in mortgage lending decisions, as discussed in [32]. An analyst is given a database of historical lending decisions and asked to ascertain if a particular demographic has been treated unfairly. The personal information about loan applicants is inherently hierarchical. For example, a person's residential address can be divided into street address, postcode, village, city, country, and so on. As historical data is often difficult to obtain, any given dataset might not include enough applicants from every fine-grained portion of the hierarchy. However, if we analysed the data at a coarser granularity, we might find a statistically significant number of participants to draw reliable conclusions. Naturally, whether hierarchical or not, demographic information is considered *highly confidential*. It is well known that even releasing summary statistics about a population can leak information about individuals in the dataset. Differential privacy has become the de facto standard for defending against such leakage. As a result, given a dataset, we wish to output its hierarchical heavy hitters privately.

*The original conference version accepted at PODS 2025 has a bug in the privacy analysis. The error was brought to our attention in personal communication with Christian Janos Lebeda, David Erb, Tudor Cebere, and Aurélien Bellet. , and details of the issue can be found in their manuscript [31].

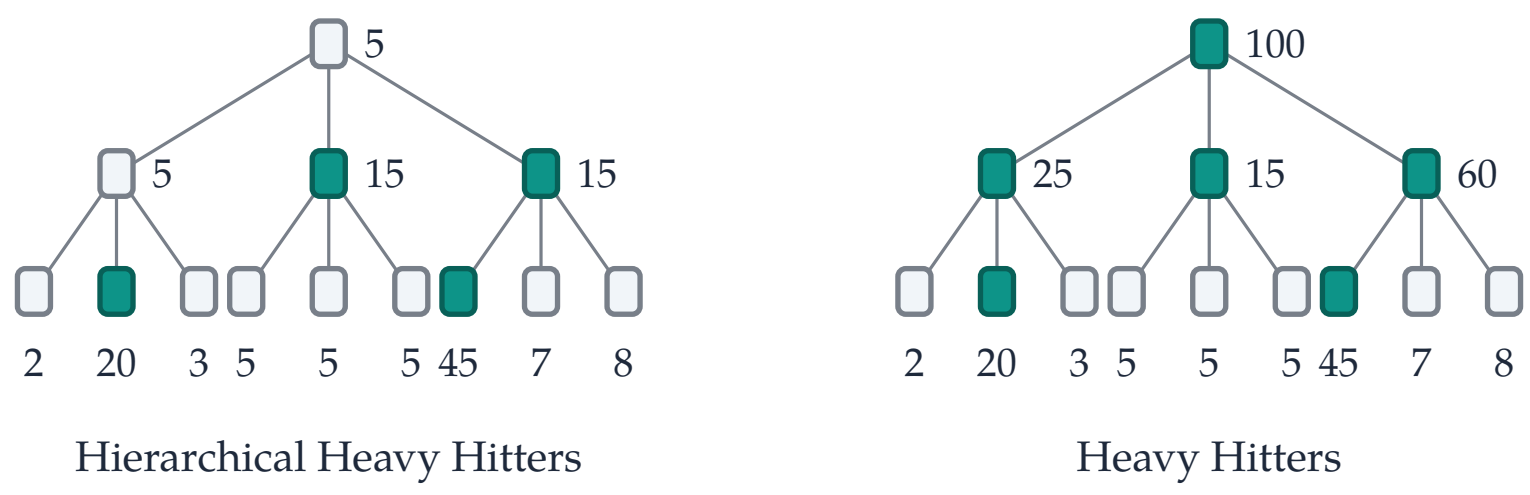

Figure 1: A dataset of 100 elements over a hierarchy with residual counts (left) and unconditional counts (right).

Note that finding hierarchical heavy hitters is *not* the same as finding heavy hitters at each level in a hierarchy (referred to as counting over trees in [23]). Hierarchical heavy hitters, which we formally define later (**Definition 2.3.5**), is a generalisation of the heavy hitters problem. At a high level, apart from telling us *if* an element is heavy, it also tells us *how* it is heavy. HHH allows us to distinguish between an element that is heavy because it has a heavy child (or a few heavy children) and an element that is heavy because it has many light children that are cumulatively heavy. Furthermore, if we are given the set of (exact) hierarchical heavy hitters of a dataset, we can derive the heavy hitters at each level of the hierarchy. The converse, however, is not true. **Figure 1** illustrates this difference with a toy example. The figure shows a dataset of 100 elements drawn from a hierarchy of height 3. Each node in the tree corresponds to an element in the hierarchical universe. The leaf nodes are fully specified elements, while the root node describes the fully generalised element of the hierarchy. The edges between nodes represent a partial order between elements of the hierarchy (see **Section 2** for formal details). Given a public threshold of $\tau = 10$, a node is heavy if its count exceeds $\tau$. The counts listed next to the nodes on the *right* tree are the underlined unconditional counts (**Definition 2.3.1**) of the node in the dataset. The nodes marked in teal on the right tree are the heavy hitters at each level of the hierarchy. The nodes marked in teal on the left tree are the hierarchical heavy hitters of the dataset. The counts listed next to the nodes on the *left* tree are called underlined residual counts (**Definition 2.3.2**). The residual count of a node is the count of a node ignoring its heavy children, whereas the unconditional count of a node is just the sum of the counts of its children. Observe that the root node is *not* a hierarchical heavy hitter, although its absolute count is greater than the threshold. The root node is heavy only because it has heavy children, not because it is an aggregation of several light children. If we were to just see the output of heavy hitters, we would lose this information.

## 1.1 Related work

### *1.1.1 Streaming HHH*

The hierarchical heavy hitter problem was first defined and studied in the streaming setting, as the offline problem is straightforward. Initial work defined the problem for streams of data drawn from a single hierarchy, and showed upper bounds on the problem, by building streaming heavy hitter summaries of data at each level [13, 33, 36]. Subsequent work extended the problem to data with multiple hierarchical attributes [11, 12], and showing lower bounds on the space required to solve the problem [26, 36]. In the streaming model, data arrives incrementally, and we assume that the algorithm does not have enough space to store the entire dataset, or enough counters for each element of the data universe. Michael Mitzenmacher, Thomas Steinke, and Justin Thaler. [36] show that approximating HHH via the Space Saving algorithm (SS) for heavy hitters [35] is optimal in terms of error and space complexity in the streaming setting.

### *1.1.2 Private Counting*

Despite its relevance to data analytics, the HHH problem has not been previously studied under differential privacy. However, there has been much research on simpler non-hierarchical heavy hitter estimation under privacy, in both the non-streaming [2, 3, 14, 24, 29], and streaming models [8, 30]. In this work, we show that despite this extensive body of work in the non-hierarchical setting, we need new algorithms to privately estimate HHH efficiently in theory and practice. The most closely related work to ours is concerned with outputting "unconditional counts" in a hierarchy, due to Ghazi et al. [23]. As any fully specified element ("leaf") affects the counts for nodes at each level of the hierarchy, we must account for it every time we release a count for a node that is an ancestor of said leaf. Hence, the DP error scales linearly with the height $h$ of the hierarchy when using basic composition, or $\sqrt{h}$ under advanced composition [20]. Alexander Edmonds, Aleksandar Nikolov, and Jonathan Ullman. [22] provide lower bounds showing that such a polynomial dependence on the hierarchy height is unavoidable if we want to estimate just the unconditional counts for every element in the hierarchy[1] with *pure* or *approximate* differential privacy (see **Appendix B**).

Badih Ghazi, Pritish Kamath, Ravi Kumar, Pasin Manurangsi, and Kewen Wu. [23] circumvent this dependence on the height of the hierarchy by relaxing the problem to consider *relative* error in estimating node counts, where the estimation gap for a node scales with the absolute count of the node[2]. Their algorithmic guarantees replace the linear dependence on the height of the hierarchy with a linear dependence of the maximum number of hierarchical heavy hitters in a dataset[3]. Although this is an asymptotic improvement, the number of heavy hitters is much larger than the hierarchy height for any practical scenario we can envisage. Hence the real-world performance will be much worse than the naive baseline of estimating the counts at each level and paying for $h$ levels of composition. Concretely, most real-world datasets are associated with shallow and wide hierarchies. Consider a dataset of bit strings of size $n = 10^6$, and a threshold of 2500, so there are up to 400 hierarchical heavy hitters. For this algorithm to improve on the simple baseline, the hierarchy would need to have more than 400 levels, implying over $2^{400}$ elements.

## 1.2 Our results

### *1.2.1 Non-Streaming Setting*

In this work, we show that the relative error[4] for any node when estimating private hierarchical heavy hitters scales by a *much* smaller constant that is independent of *both* the height of the hierarchy and the number of hierarchical heavy hitters in the tree. Our algorithm is simpler to describe than that of [23:Algorithm 1], and it can be used to solve the more general problem with better error guarantees than prior attempts would imply. More specifically, the variance parameter of the noise added to each element is *independent* of *both* the height of the hierarchy and the number of heavy hitters. At a high level, an intuitive explanation is that by targeting relative error instead of absolute error, elements higher up in the hierarchy with larger frequencies can tolerate more DP noise. This structure proves to be critical for circumventing composition bounds, by allowing us to re-use information about lower regions of the hierarchy, and apply them to higher regions of the hierarchy. Bounding the absolute error for every node requires us to treat each node independently, therefore destroying the structure we leverage to propose more accurate algorithms.

### *1.2.2 Streaming Setting*

In the streaming setting, along with DP error and composition error, we also need to account for the approximation error due to space constraints. The main issue with streaming algorithms is that although the exact version of the function (exact hierarchical heavy hitters) has low global sensitivity (as counting queries are 1-sensitive), the approximation function can have high global sensitivity. For instance, the Space Saving (SS) algorithm described by Michael Mitzenmacher, Thomas Steinke, and Justin Thaler. [36] is optimal in the non-private setting, but T -H Hubert Chan, Mingfei Li, Elaine Shi, and Wenchang Xu. [8] show that the global sensitivity of the approximation function induced by SS scales linearly with the number of counters per sketch (denoted with $\kappa$ in this document). This would imply that scale of the noise used per node also scales linearly with $\kappa$. However, Christian Janos Lebeda and Jakub Tetek. [30] improve on this naive bound by providing a DP mechanism for non-hierarchical heavy hitters, where the estimation error of the protocol does not rely on global sensitivity. Inspired by [30], we design algorithms for hierarchical heavy hitters with DP noise whose variance is independent of $\kappa$. The intuition behind our algorithm is that although the approximation algorithm we use has high sensitivity, the counters in a sketch are highly correlated, with few degrees of freedom. This structure (correlated counters) can be used to bypass composition bounds typically enforced due to high global sensitivity of the function. This observation is closely related to why the seminal Sparse Vector Technique algorithm (SVT) by Cynthia Dwork, Moni Naor, Omer Reingold, Guy N Rothblum, and Salil Vadhan. [19] also circumvents composition bounds. Although the two appear unrelated at first glance, we show that releasing private counts for our sketching algorithm and SVT algorithm are essentially equivalent and use the *same* underlying theoretical concepts to bypass composition.

[1] This is a strictly simpler problem than HHH, which involves estimating unconditional counts. Therefore these lower bounds immediately apply to HHH as well.

[2] In other words, nodes with large unconditional frequencies are allowed to tolerate more estimation error than nodes with smaller unconditional frequencies.

[3] Algorithm 1 of their paper uses the constant $c$, independent of the height of the hierarchy, as an upper bound on the maximum number of hierarchical heavy hitters.

[4] We use a slightly different definition of relative error than [23], but one which we believe is more natural. Nevertheless the argument holds in general.

More broadly, the message of this work is that *"where we have structure (sparsity, monotonicity, correlation, etc.), we can leverage this structure to circumvent basic or advanced composition bounds"*. Hierarchies offer structure in that the frequency of elements higher up in the hierarchy is computed using frequencies of elements lower down. In streams, we show that despite high global sensitivity, we can leverage correlation between counters in a sketch to circumvent composition bounds. We refer the reader to **Appendix A** for further discussion on the role of structure in circumventing composition. To summarize:

1. In **Section 3**, we propose the first known private algorithm for the task of hierarchical heavy hitter estimation. In the non-streaming setting, the relative error of our algorithm is independent of the height of the hierarchy and the number of hierarchical heavy hitters in the dataset.
2. In the streaming setting (**Section 4**), we show that the DP error of our HHH estimation algorithm is independent of the space bound. However, in this setting, the DP error still depends on the height of the hierarchy. Therefore, there is a gap in relative estimation error between the streaming setting and the non-streaming setting under privacy. Despite this gap, for all practical situations, removing the dependence on space is far more critical than the dependence on the height of the hierarchy (as the number of counters is often orders of magnitude larger than the height of the hierarchy).

The rest of the paper is organised as follows. In **Section 2**, we formally introduce the problem of hierarchical heavy hitter estimation and review preliminary results from differential privacy. In **Section 3** we describe our solution in the non-streaming setting with unlimited space. In **Section 4**, we describe our algorithm in the streaming setting.

## 2 Prelims and Problem Statement

### 2.1 General Notation

We describe sets with calligraphic font $\mathcal{H}$. For a probability distribution $D$, we denote with $x \leftarrow\!\!\$\, D$ the event of sampling $x$ according to $D$. We highlight random variables and samples to distinguish them from constants, as shown above. For a randomized algorithm $A$, the notation $A(X; z, w)$ means that $A$ is run on input $X$ with random samples $z$ and $w$. If an argument is written without highlighting, as in $A(X; z, w)$, that value is fixed. For example, $A(X; \vec{w}, \vec{v})$ conditions on the fixed value $\vec{w}$ and leaves only $\vec{v}$ random. We write $[n]$ to denote the set $\{1, ..., n\}$. For any event $E$, we denote with $\overline{E}$, the complement of the event. Let $\vec{z} = (z_i)_{i \in \mathcal{H}}$ be a vector indexed by an ordered set $\mathcal{H}$. Given ordered subset $\mathcal{J} = (j_1, ..., j_m) \subseteq \mathcal{H}$, the *restriction* of $\vec{z}$ to $\mathcal{J}$ is given by

$$\vec{z}_{\mathcal{J}} := \left(z_{j_1}, ..., z_{j_m}\right). \tag{1}$$

When an algorithm $A$ has vector-valued output, we write $A(X; \vec{w}, \vec{v})_{\mathcal{J}}$ for that output restricted to $\mathcal{J}$. Next, we first review the necessary tools from differential privacy and its basic properties, and then formalise the idea of a hierarchical domain.

### 2.2 Differential Privacy Definitions

We begin the standard definition of neighbouring datasets.

Let $X$ and $X'$ be a multi sets of elements picked from some (possibly hierarchical) domain $\mathcal{H}$. $X$ and $X'$ are said to be neighbouring, (denoted as $X \sim X'$), if they differ by one element only, i.e., $X' = X \cup \{x'\}$, or vice-versa. The following definition of differential privacy is often referred to as *addition & removal* differential privacy[5].

**Definition 2.2.1 (Differential Privacy (DP))** Fix some function $f$ that maps a set of elements from a hierarchical domain $\mathcal{H}$ to some range $\mathcal{Y}$. Fix $n \in \mathbb{N}$. Let $X \in \mathcal{H}^n$ and $X' \in \mathcal{H}^{n+1}$ denote *any* pair of neighbouring datasets. For $\varepsilon > 0$ and $\delta = \mathsf{negl}(n)$, where $\mathsf{negl}(\cdot)$ is a negligible function in $n$, we say a random algorithm $M$ computes $f$ with $(\varepsilon, \delta)$-differential privacy *if and only if* for *all* $\mathcal{A} \subseteq \mathcal{Y}$,

$$e^{-\varepsilon}(\Pr[M(X', f) \in \mathcal{A}] - \delta) \le \Pr[M(X, f) \in \mathcal{A}] \le e^{\varepsilon} \Pr[M(X', f) \in \mathcal{A}] + \delta$$

The special case of $(\varepsilon, \delta)$-DP with $\delta = 0$ is referred to as pure DP, whereas $\delta > 0$ is known as approximate DP. A standard approach to obtain DP is to add noise proportional to the global sensitivity of the function being evaluated. Given any function $f$ that maps a set of elements from a hierarchical domain $\mathcal{H}$ to $\mathbb{R}$, we define the global sensitivity $\Delta_G(f)$ of $f$ as

$$\Delta_G(f) := \max_{(X, X')} \|f(X) - f(X')\|_1, \tag{2}$$

where the maximum is taken over *any* pair of neighbouring datasets $X, X'$.

It is well known that the global sensitivity of a counting query, such as the queries defined in **Definition 2.3.1** and **Definition 2.3.2** is 1. The following facts about differential privacy can be found in any introductory textbook on differential privacy [20].

**Theorem 2.2.2 (Laplace Histograms )** Let $f$ be a function that maps elements from some domain to a subset of $\mathbb{R}^d$, that has global sensitivity $\Delta_G$. Then the Laplace mechanism $M$ defined as $M(X) = f(X) + (Y_1, ..., Y_d)$ where each $Y_1, ..., Y_d \leftarrow \mathsf{Laplace}(\Delta_G/\varepsilon)$ is $\varepsilon$-DP.

[5] See **https://differentialprivacy.org/one-sided/** for a more complete description of the nomenclature. The constants in the relative error of our results for non-streaming HHH can be further improved if we focused on just *removal* only privacy, but we choose to tackle the more complete definition.

**Theorem 2.2.3 (Basic Composition )** Let $M_1$ and $M_2$ be a $(\varepsilon_1, \delta_1)$-DP and $(\varepsilon_2, \delta_2)$-DP algorithm respectively. $M(X) = (M_1(X), M_2(X))$ is $((\varepsilon_1 + \varepsilon_2), (\delta_1 + \delta_2))$-DP.

**Theorem 2.2.4 (Post Processing)** Let $A_1$ be an $(\varepsilon, \delta)$-DP algorithm and $A_2$ be a (possibly randomised) post-processing algorithm. Then the algorithm $A(x) = A_2(A_1(x))$ is still an $(\varepsilon, \delta)$-DP algorithm.

## 2.3 Hierarchies

Formally, a hierarchical domain is a set $\mathcal{U}$ associated with a partial order $(\succ)$. In streaming literature [13, 36], this partial order is often represented by a function called $\mathsf{Generalise} : \mathcal{U} \to \mathcal{U}$ which maps elements of the universe to other elements in the universe[6]. In this work, it suffices to think of a hierarchical domain as the set of elements that has one to one mapping with the nodes of a rooted tree with finite arity[7]. For any element $x \in \mathcal{U}$, $\mathsf{Generalise}(x)$ refers to the parent or prefix of $x$. We say an element $*$ is *fully generalised* or the root of the tree if $\mathsf{Generalise}(*) = *$. We say an element $e$ is *fully specified* if there exists no $s \in \mathcal{U}$ such that $e = \mathsf{Generalise}(s)$ ($e$ is a leaf node of the rooted tree representing the hierarchy). We denote by $\mathsf{Generalise}^{(k)}(x)$ as the ancestor of $x$ that is $k$ steps away in the tree representation (element obtained by applying $\mathsf{Generalise}$ $k$ times on $x$). The pair $(\mathcal{U}, \mathsf{Generalise})$ defines a hierarchical domain $\mathcal{H}$. The height $h$ of the hierarchy represents the maximum number of times *any* fully specified element must be generalised to get a fully generalised element (or more simply, the height of the tree representing the hierarchy). As a concrete example of a single-dimensional hierarchy, one can imagine $\mathcal{U}$ to be the set of prefixes of $h$-bit strings. When $h = 4$, this universe has 16 fully specified elements: 0000, 0001, ..., 1111. The prefix $000*$ is a generalisation of 0000 and 0001. We use notation $x \succ p$ (read as $p$ is reachable from $x$) if there exists a $k \in \mathbb{N}$ such that $p = \mathsf{Generalise}^{(k)}(x)$, and $x \succeq p$ if $p = \mathsf{Generalise}^{(k)}(x) \vee x = p$. For any $p \in \mathcal{H}$, $\mathsf{Succ}(p) := \{q \in \mathcal{H} : q \succ p\}$ denotes the strict successors of $p$, and $\mathsf{Par}(p) := \{q \in \mathcal{H} : p \succeq q\}$ the ancestors of $p$ (including $p$ itself). Henceforth, we assume that any dataset $X$ is a multiset of fully specified elements from some hierarchical universe $\mathcal{H}$ of height $h$. Throughout, we use the the convention wehere $X' := X \cup \{x'\}$ is the larger multiset and has the extra fully specified element $x'$.

**Definition 2.3.1 (Unconditional Count or Absolute Frequency)** Given a dataset $X$, the unconditional frequency of any element $p \in \mathcal{H}$, denoted by $f_X(p)$, is the number of elements in $X$ that generalise to $p$.

$$f_X(p) = \sum_{e \in X} \mathbb{1}[e \succeq p]$$

In **Figure 1**, the unconditional counts of each node are written next to each node in the tree for the figure on the *right*.

**Definition 2.3.2 (Residual Count / Conditional Frequencies)** Given a dataset $X$, and a set $\mathcal{S} \subseteq \mathcal{H}$, we say $x \not\succeq \mathcal{S}$ if there does not exist $q \in \mathcal{S}$ such that $x \succeq q$. We define the conditional or residual count $F_{X,\mathcal{S}}(p)$ of a prefix $p$ with respect to $\mathcal{S}$ as the sum of all fully specified elements who do not have a parent already in $\mathcal{S}$.

[6] $\mathsf{Generalise}$ encodes partial order binary relation $R : \mathcal{U} \times \mathcal{U} \to \{0, 1\}$, such that $R(x, p) = 1 \iff p = \mathsf{Generalise}(x)$

[7] The HHH literature also considers multi-dimensional hierarchies, where the universe is represented by nodes of a lattice rather than a rooted tree. However, in this work, we focus on single dimensional hierarchies which can be represented by a tree.

$$F_{X,\mathcal{S}}(p) := \sum_{e \in X \wedge\, e \succeq p \wedge\, e \not\succeq \mathcal{S}} f_X(e)$$

We will use $F_{X',\mathcal{S}'}(p) := \sum_{e \in X \wedge\, e \succeq p \wedge\, e \not\succeq \mathcal{S}} f_{X'}(e)$ to denote the residual count when using $X' := X \cup \{x'\}$ as the input dataset. In the left *tree* in **Figure 1**, the set $\mathcal{S}$ is shown by nodes in teal, and the conditional count with respect to $\mathcal{S}$ is written by each node.

**Definition 2.3.3 (Level Of A Node / Prefix)** The level of a prefix $p \in \mathcal{H}$ is the (minimum) number of applications of Generalise to reach $*$ i.e. $\mathsf{Level}(p) = k \iff * = \mathsf{Generalise}^{(k)}(p)$.

For example, let $\mathcal{H}$ be the set of 4-bit bistrings. If we generalise $011*$ three times we get to $*$, so the level of the prefix is 3. The fully specified elements or leaf elements are at level $h = 4$ and $*$ is at level 0. With these definitions in place, we can formally define the concept of a heavy hitter and a hierarchical heavy hitter.

**Definition 2.3.4 (Exact Heavy Hitters)** For dataset $X$ and threshold $\tau \in \mathbb{R}$, we say prefix $p$ is a heavy hitter (HH) if $f_X(p) > \tau$. The set of heavy hitters of $X$ is $\mathcal{HH} = \{e \in X : f_X(e) \geq \tau\}$.

**Definition 2.3.5 (Exact Hierarchical Heavy Hitters)** The set of exact hierarchical heavy hitters is defined inductively. Let $X$ denote a dataset drawn from a hierarchy of height $h$, then

1. $\mathcal{HHH}_h$ denotes the exact heavy hitters in $X$.
2. For any prefix $p$ at level $0 \leq l < h$, let $F_{\mathcal{HHH}_{l+1}}(p)$ be the residual count (Defn. Definition 2.3.2) of $p$ given $\mathcal{HHH}_{l+1}$. Then $\mathcal{HHH}_l$ is defined as $\mathcal{HHH}_{l+1} \cup \left\{p \in \mathsf{Level}(l) : F_{\mathcal{HHH}_{l+1}}(p) \geq \tau\right\}$.
3. $\mathcal{HHH}_0$ is the set of *exact* hierarchical heavy hitters of $X$.

As discussed earlier, **Figure 1** illustrates this difference between heavy hitters and hierarchical heavy hitters.

### *2.3.1 Approximate Hierarchical Heavy Hitters*

In this paper, the function $f$ that our algorithm $M$ computes is the hierarchical heavy hitters of a dataset $X$. By definition, differential privacy restricts us from outputting exact answers or using a deterministic algorithm to compute approximate values. As the output *must* be random, we can no longer output exact counts of the hierarchical heavy hitter problem. Thus, keeping in line with the definitions introduced in [13, 36] we define the task of *approximate heavy hitters*, where the estimates are within some approximation error $\alpha$ with high confidence $1 - \beta$. As we now have noise in the system, we relax the threshold by $\alpha$ units, where $\alpha$ allows the error to grow larger for larger values (i.e., relative error). Clearly the smaller the value of $\alpha$, the closer we are to the definition of exact hierarchical heavy hitters. The coverage constraint says we relax our definition of what is heavy due to DP noise i.e., prevent false positives. Our goal is to come up with a theoretical bound on the error, and show that the error is small enough for practical use cases.

**Definition 2.3.6 (Private Approximate Hierarchical Heavy Hitters)** Let $M$ denote an algorithm that receives as input a multi set $X$ of $n$ fully specified elements from some hierarchical domain $\mathcal{H}$. Fix a public threshold $\tau \in \mathbb{R}$, a confidence parameter $\beta \in (0, 1)$, privacy parameters $\varepsilon \in (0, \log n)$ and $\delta = \mathsf{negl}(n)$. We say the algorithm $M$ correctly finds approximate private hierarchical heavy hitters with relative error $(\beta, \alpha)$ if it outputs $\mathcal{S} \subseteq \mathcal{H}$ and approximate counts $\tilde{f}_X(p)$ such that:

1. **Privacy**: $M$ is $(\varepsilon, \delta)$-DP.
2. **Simultaneous Relative Error**: With probability $1-\beta$ we have $\max_{p \in \mathcal{H}} \left| \frac{f_X(p) - \tilde{f}_X(p)}{f_X(p)} \right| \leq O(\alpha)$.
3. **Coverage**: With probability $1-\beta$, every $p \in \mathcal{S}$ satisfies $F_{X,\mathcal{S}}(p) \geq \tau - \alpha$ (no false positives), and every $p \notin \mathcal{S}$ satisfies $F_{X,\mathcal{S}}(p) \leq \tau + \alpha$ (with $F_{X,\mathcal{S}}(p)$ as defined in **Definition 2.3.2**).

We want to show that in the non-streaming setting, the relative error $\alpha$ of our algorithm does not grow linearly with the the height of the hierarchy[8], and is independent of the number of heavy hitters in the hierarchy (as discussed in the introduction above). In the streaming setting we will have to deal with error due to privacy and lack of space. Thus, the streaming version of our problem is the exact same problem with limited space.

**Definition 2.3.7 (Streaming Private HHH)** The streaming problem is to solve **Definition 2.3.6** with a constant amount of space $\kappa = O(1)$.

Of course, to prevent the problem from being degenerate, we will assume that the amount of space available is significantly smaller than the size of the stream $n$ or the size of the universe $|\mathcal{H}|$.

## 3 Offline Private Hierarchical Heavy Hitters

In this section, we tackle the offline version of the problem, with no space constraints. Before describing the complete protocol and showing how it solves **Definition 2.3.6**, we provide an intuitive explanation of the techniques used to circumvent the dependence the height of the hierarchy and the number of hierarchical heavy hitters by working through a sequence of attempts. Let $\mathcal{H}$ denote a hierarchy with height $h$ and let $X$ denote a dataset of $n$ fully specified elements drawn from $\mathcal{H}$.

### 3.1 Laplace Histograms

Observe that the unconditional frequencies for each prefix of the hierarchy can be estimated by computing $h$ histograms for each level of the hierarchy. For two neighbouring datasets $X$ and $X'$ that differ by a single input only, there is *exactly* one node per level for which the unconditional frequency of the nodes differ by one under $X$ and $X'$. Thus, a first approach to hierarchical heavy hitter estimation is to release $h$ Laplace histograms by adding noise drawn from a Laplace noise distribution with scale $\frac{h}{\varepsilon}$ to every node in the hierarchy. By the privacy of the Laplace histograms (**Theorem 2.2.2**), each level is $\varepsilon/h$-private. As any node contributes to at most $h$ counts, by basic composition and the privacy of the Laplace mechanism, the set of realised counts is $(\varepsilon, 0)$-DP. Then, one could compute hierarchical heavy hitters via post-processing, which does not affect the privacy of the algorithm (**Theorem 2.2.4**). As we add noise with scale $\frac{h}{\varepsilon}$ to each node in $\mathcal{H}$, the DP error per node scales linearly in the height of the hierarchy. Furthermore, if we wanted to upper bound the simultaneous error over all prefixes in the hierarchy, then we need to apply the union bound over all nodes of the hierarchy. As the number of nodes is exponential in the height of the hierarchy, the relative estimation error scales $\mathsf{poly}(h)$.

[8] As we want to bound the worst case simultaneous relative error for *any* element in the hierarchy, logarithmic dependence on height is unavoidable (via the union bound).

### 3.2 Stability Histograms

One solution to circumventing such a union bound over all the elements of the universe when the size of the universe is very large is to use stability histograms [2, 6, 42] instead of Laplace histograms. For fixed privacy parameters $\varepsilon \in (0, \log n)$ and $\delta = \mathsf{negl}(n)$, the key intuition behind stability histograms is that if we only released private estimates (using the Laplace mechanism) for nodes with *large enough* non-zero frequency, the simultaneous error bound improves from $\log(|\mathcal{H}|)$ to $\log(n) < \log(1/\delta)$ (as there are at most $n$ elements in $X$). This is advantageous when $|\mathcal{H}| \gg n$ meaning that many nodes have zero frequency. For such nodes, we incur no error at all (as the algorithm ignores them).

However, this technique cannot achieve pure DP as was possible in the Laplace histogram case. Observe that if $X'$ contains an element $x'$ that is not present in $X$ (we call such an $x'$ isolated), an adversary can perfectly distinguish between $X$ and $X'$ if the count of this element, albeit noisy, appeared in the output (as when processing $X$, nodes representing generalisations of $x'$ would have frequency $0$, and would be ignored). Nevertheless, since stability histograms only output counts for elements with "large" counts, and as the frequency of such an isolated element $x'$ is $1$ (from the definition of neighbouring datasets), we can set the frequency threshold for "large enough" such that with probability at least $1 - \delta$ the isolated element will never show up in the final output. This is sufficient to achieve $(\varepsilon, \delta)$-DP. Over the whole tree representing the hierarchy, the case where $x'$ distinguishes $X$ and $X'$ can occur at at most $h$ levels of $\mathcal{H}$. Thus, if we output stability histograms with $\delta' = \delta/h$ at each level then, by basic composition, the final output does not contain any isolated elements with probability at least $1 - \delta$. In this way, stability histograms allow us to circumvent the union bound over all the nodes in $\mathcal{H}$, at the cost of going from pure to approximate DP. It does not however, circumvent the issue discussed above where the scale of the DP noise is $\frac{h}{\varepsilon}$.

In recent work, Christian Janos Lebeda, David Erb, Tudor Cebere, and Aurélien Bellet. [31] apply stability histograms to finding a set of heavy nodes within a hierarchy. Their algorithm performs a set of parallel binary searches up and down the hierarchy, ensuring that each input element is only queried $\log h$ times, so that the scale of the DP noise is poly-logarithmic in $h$ instead of polynomial in $h$. Their notion of "heavy node" is similar to that of hierarchical heavy hitters, but there is no concept of residual count: any node marked as heavy automatically implies that all of its ancestors are heavy as well. Thus, this stability histogram-based approach does not solve the DP-HHH problem, and suggests a separation between the two problems of finding heavy nodes in a hierarchy and of finding hierarchical heavy hitters.

### 3.3 DP Counting On Trees

One approach to resolving this issue is to make repeated use of the Above-Threshold Algorithm by Cynthia Dwork, Moni Naor, Omer Reingold, Guy N Rothblum, and Salil Vadhan. [19]. Given a hierarchy $\mathcal{H}$ then, by definition, there can be at most $c = \frac{n}{\tau - \Delta}$ hierarchical heavy hitters. Observe that $c$ is a constant that depends only on $\tau$ and $\Delta$, and is independent of the height of the hierarchy $h$. When the height of the hierarchy $h$ is large and $c \ll h$, we would incur lower per-node DP error if the scale of the noise were $\frac{c}{\varepsilon}$ instead of $\frac{h}{\varepsilon}$. Badih Ghazi, Pritish Kamath, Ravi Kumar, Pasin Manurangsi, and Kewen Wu. [23] observed this fact and proposed an algorithm that traverses the tree bottom up. At each level of the tree, their algorithm inspects only one node, the one with maximal unconditional count for that level. There are at most $h$ such nodes (one per level). However, even in the worst case, at most $c$ out of $h$ of these nodes can be conditionally heavy. Thus, we have $h$ counting queries, out of which $c$ are conditionally heavy. Their solution implicitly uses the Sparse Vector Technique (SVT) by composing Above-threshold $h$ times, but they pay only for the $c$ conditionally heavy nodes. Once $c$ levels have been identified, the algorithm prunes the tree, and restricts it to just the $c$ levels with at least one conditionally heavy node. Then it estimates the counts of the pruned tree by solving $c$ instances of the private stability histogram estimation problem[9] with privacy budget $\varepsilon/c$. As commented in the introduction, although this is an asymptotic improvement, it is very hard to imagine cases where $c \ll h$. Thus, the simple Laplace stability histograms with composition error $h$ would outperform the algorithm by Badih Ghazi, Pritish Kamath, Ravi Kumar, Pasin Manurangsi, and Kewen Wu. [23] in almost all feasible situations.

### 3.4 Our Approach

Our algorithm is based on the following critical observation. Although it is true that there are at most $c$ conditionally heavy levels in the tree, *only one* of the conditionally heavy nodes is truly influenced by $x'$. For the remainder of the heavy nodes, there is no difference in the conditional frequencies of $X$ and $X'$. It is this structure that we want to exploit to improve error our guarantees. In prior work, Haim Kaplan, Yishay Mansour, and Uri Stemmer. [28] show that there are instances where the composition error of the SVT algorithm is suboptimal, even when dealing with a stream of adaptively chosen queries by an unbounded adversary. Instead, they propose a general algorithm (called Threshold Monitor) where DP error of a query scales linearly by a constant $O\left(\frac{\varepsilon(k+1)}{\log 1/\delta}\right)$, where $k \geq 1$ represents the number of times any data point can contribute to a counting query being heavy. In this work, we revisit the Threshold monitor algorithm under a simplified regime. We note that our setting has a few advantages. Instead of adaptively chosen arbitrary counting queries, we will have a *pre-determined set of monotonically increasing counting queries*. Additionally for us, the structure of our problems corresponds to the case $k = 1$, which can simplify the analysis. In estimating HHH in the non-streaming setting with central differential privacy, we can pre-select the order of queries to be bottom up (leveraging monotonicity and limited influence of any node), and we do not have to deal with adaptive queries. Thus, in the case of hierarchical heavy hitter estimation, paying a price of $c/\varepsilon$ of SVT or the larger constants of Threshold Monitor for adaptivity in composition is wasteful[10]. Instead, we use the general idea of threshold monitor, and with more direct privacy analysis we show that DP noise per node scales $c_0/\varepsilon$ instead of $c/\varepsilon$, where $c_0$ is truly a constant since it is independent of *both* the height of the hierarchy and the number of heavy hitters in $\mathcal{H}$. Furthermore it is smaller than the constants used in threshold monitor. **Algorithm 1** shows our algorithm for the non-streaming case.

[9]While Badih Ghazi, Pritish Kamath, Ravi Kumar, Pasin Manurangsi, and Kewen Wu. do not explicitly refer to stability histograms, their use of bounded truncated Laplace noise is equivalent. In the case of bounded noise, the union bound is affected by the size of the support of the truncated noise distribution, which is upper bounded by $O(1/\delta)$. For stability histograms, it is affected by the number of elements which is upper bounded by $1/\delta$.

[10]In the proofs we will pin-point exactly which parts of the general Threshold monitor algorithm we do not need, and instead can use simpler claims which led to improved constants.

**Algorithm 1:** Non-Streaming DP-HHH Detection

**Input:** Data $X$ of size $n$ over hierarchy $\mathcal{H}$ with height $h$, Privacy parameter $\varepsilon \in (0, \log n)$, $\delta = \mathsf{negl}(n)$, Threshold $\tau > 0$, Confidence parameter $\beta \in (0, 1)$

```
1   c_0 := log_{5/4}(1/δ)
2   η := ε / log(1/δ)
3   Δ := (1/η) log(1/η) ≥ 1
4   ξ := 2ηc_0
5   If τ < 24Δ log(2h/(δβ)), then
6   |  output {} and end program as threshold is too low for stability histograms.
7   for i = h, ..., 1 do
8   |  A_i = {p ∈ H | Level(p) = i}
9   |  for p ∈ A_i do
10  |  |  If F_{X,S}(p) = 0 then
11  |  |  |  Continue, ignoring this node.
12  |  |  w_p ←$ Lap(6Δ)
13  |  |  v_p ←$ Lap(1/η)
14  |  |  v̄_p := min{v_p, Δ}
15  |  |  if F_{X,S}(p) + w_p + v̄_p ≥ τ then
16  |  |  |  Output b_p = ⊤
17  |  |  |  Update S = S ∪ {p}
18  |  |  |  Output F̃_{X,S}(p) = F_{X,S}(p) + Lap(ξ)
19  |  |  end if
20  |  end for
21  end for
22  Output S and {F̃_{X,S}(p)}_{p∈S}.
```

Algorithm 1: Non-Streaming DP-HHH Detection

Before stating our theorem statement, we introduce notation that is used throughout this section. The above algorithm can be broken down into two steps. First output the set $\mathcal{S}$ with the tops and bots. Then, use the Laplace mechanism to release counts. The main technical challenge will be to prove that the release of $\mathcal{S}$ is private. Much like the original SVT analysis, we can just use sequential composition of the release of actual counts to argue full privacy. Observe that once we fix the output of the algorithm to $\vec{b} \in \{\perp, \top\}^{|\mathcal{H}|}$, this fully determines the set of hierarchical heavy hitters $\mathcal{S}$. That is,

$$\mathcal{S} := \mathcal{S}\left(\vec{b}\right) = \left\{p \in \mathcal{H} : b_p = \top\right\} \tag{3}$$

Under the same output $\vec{b}$, the neighbour $X' = X \cup \{x'\}$ induces the same selected set, $\mathcal{S}' := \mathcal{S}'\left(\vec{b}\right) = \mathcal{S}$. Next observe, that given $\vec{b}$ we can partition $\mathcal{H}$ into 3 sets $\mathcal{I}_{\text{Active}}, \mathcal{I}_{\text{Unrelated}}, \mathcal{I}_{\text{After}}$ as shown in **Figure 2**.

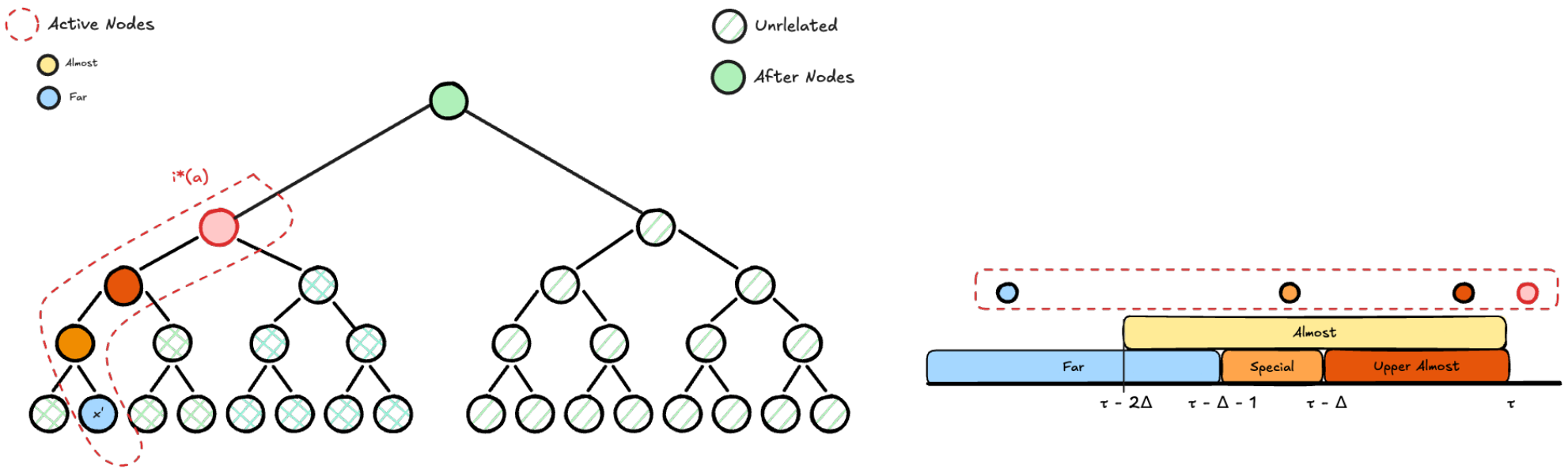

Figure 2: Given any outout $\vec{b} \in \{\perp, \top\}^{|\mathcal{H}|}$ we can partition $\mathcal{H}$ into 3 sets $\mathcal{I}_{\text{Active}}, \mathcal{I}_{\text{Unrelated}}, \mathcal{I}_{\text{After}}$. The key insight is that $x'$ does not affect the counts of nodes in $\mathcal{I}_{\text{Unrelated}}$ so these nodes will not affect privacy analysis. The nodes in $\mathcal{I}_{\text{After}}$ have the *same* residual counts in $X$ and $X'$ as $x$'s contribution has been removed at this point. It gets removed exactly at $i^\star$ marked in red solid colour inside the dotted block.

Before we can formally define the conditions of membership of the above sets, we need to define the Inflexion operator.

**Definition 3.4.1 (Inflexion Operator And The Stopped Path)** Fix a sequence of counting queries $(p_1, \ldots, p_h)$ ordered from leaf-to-root and an output vector $\vec{a} \in (\perp, \top)^h$. We define the inflexion operator by

$$\mathsf{Inflexion}(\vec{a}) := \min\left\{i \in [h] : a_{p_i} = \top\right\}.$$

If no such index exists, set $\mathsf{Inflexion}(\vec{a}) = h + 1$. We also define the stopped path induced by $\vec{a}$ as

$$\mathcal{P}(\vec{a}) := \left(p_1, \ldots, p_{\min\{\mathsf{Inflexion}(\vec{a}), h\}}\right).$$

Although defined for any path, throughout this work the path we focus on is the leaf-to-root path containing $x'$. To prevent notation pollution, we use $i^\star(\vec{a}) := \mathsf{Inflexion}(\vec{a})$ as short hand for the inflexion operator. Let $\vec{a} = \left(b_{p_1}, \ldots, b_{p_h}\right)$ denote the restriction of the full output $\vec{b}$ to the leaf-to-root path involving $x'$,

Then

$$\mathcal{I}_{\text{Unrelated}} := \{p \in \mathcal{H} : x' \not\preceq p\} \tag{4.1}$$

$$\mathcal{I}_{\text{Active}} := \left\{p_1, \ldots, p_{\min\{\mathsf{Inflexion}(\vec{a}), h\}}\right\} \tag{4.2}$$

$$\mathcal{I}_{\text{After}} := \mathcal{H} \setminus (\mathcal{I}_{\text{Active}} \cup \mathcal{I}_{\text{Unrelated}}) \tag{4.3}$$

In **Figure 2**, the set of nodes in $\mathcal{I}_{\text{Active}}$ is denoted by red dotted lines, $\mathcal{I}_{\text{Unrelated}}$ are marked with green stripes and $\mathcal{I}_{\text{After}}$ is marked in green solid colour.

> **Remark:** We draw the readers attention to a seemingly obvious fact, but one that is easily misunderstood when reading the analysis later with all the moving parts. The location of $i^\star$ is a function of the output stream restriction $\vec{a}$ *only*. The internal randomness of **Algorithm 1** determines the probability of seeing said output $\vec{a}$, but the location $i^\star$ is determined by the output and the output alone.

Now with the above partition defined, we can further classify the nodes in $\mathcal{I}_{\text{Active}}$ using the samples $\vec{w}$ drawn by **Algorithm 1**.

**Definition 3.4.2 (Sets Induced By Random Samples And Output)** Let $\Delta \geq 1$ be as defined in **Algorithm 1**. Fix any full output vector $\vec{b} \in (\perp, \top)^{|\mathcal{H}|}$ and path $(p_1, ..., p_h)$ ordered from leaf to root. Let $\vec{a} := \left(b_{p_1}, ..., b_{p_h}\right) \in (\perp, \top)^h$ be the restriction of $\vec{b}$ to this path. Let $\mathcal{P} := \mathcal{P}(\vec{a})$ be the stopped path from **Definition 3.4.1**. For any vector $\vec{w} = \left(w_p\right)_{p \in \mathcal{P}} \in \mathbb{R}^{|\mathcal{P}|}$, define the following (overlapping) subsets of $\mathcal{P}$.

$$\mathcal{I}_{\text{Inflex}}(\vec{a}) := \left\{p \in \mathcal{P} : a_p = \top\right\}$$

$$\mathcal{I}_{\text{Far}}(\vec{w}, \vec{a}) := \left\{p \in \mathcal{P} : a_p = \perp \wedge F_{X,\mathcal{S}}(p) + w_p < \tau - \Delta - 1\right\}$$

$$\mathcal{I}_{\text{Almost}}(\vec{w}, \vec{a}) := \left\{p \in \mathcal{P} : a_p = \perp \wedge F_{X,\mathcal{S}}(p) + w_p \geq \tau - 2\Delta\right\}$$

We also use the following subsets of $\mathcal{I}_{\text{Almost}}(\vec{w}, \vec{a})$.

$$\mathcal{I}_{\text{Special}}(\vec{w}, \vec{a}) := \left\{p \in \mathcal{I}_{\text{Almost}}(\vec{w}, \vec{a}) : \tau - \Delta - 1 \leq F_{X,\mathcal{S}}(p) + w_p < \tau - \Delta\right\}$$

$$\mathcal{I}_{\text{Upper-Almost}}(\vec{w}, \vec{a}) := \left\{p \in \mathcal{I}_{\text{Almost}}(\vec{w}, \vec{a}) : F_{X,\mathcal{S}}(p) + w_p \geq \tau - \Delta\right\}$$

Once again **Definition 3.4.2** is defined generally for any leaf to root path, but for our analysis it will always be the path containing $x'$. The nodes from $\mathcal{I}_{\text{Active}}$ can be projected onto the real number line as shown on the right in **Figure 2** and that is the picture to think of whenever we write $\mathcal{P}(\vec{a})$ downstream.

> **Remark:** As mentioned earlier, $\mathcal{I}_{\text{Active}}$ is determined by $\vec{a}$ which is a restriction of the full output $\vec{b}$, to just the leaf-to-root path starting from leaf that contains $x'$. Once $\mathcal{I}_{\text{Active}}$ is defined the subsequent subsets $\mathcal{I}_{\text{Far}}, \mathcal{I}_{\text{Special}}, \mathcal{I}_{\text{Upper-Almost}}$ together with the inflexion node $\mathcal{I}_{\text{Inflex}}$ partition the stopped path $\mathcal{I}_{\text{Active}}$; the first three depend *only* on the samples $\vec{w}$ (given $\vec{a}$) and *not* on the samples $\vec{v}$.

Intuitively, what we are saying is that although $\mathcal{I}_{\text{Active}}$ could have size $h$, only certain nodes will be "dangerous", and contribute to our privacy budget. These are the nodes that correspond to queries in $\mathcal{I}_{\text{Special}}$ and $\mathcal{I}_{\text{Upper-Almost}}$, as these are such that the value of $\vec{v}$ could allow us to distinguish between $X$ and $X'$. As we have $\overline{v} \leq \Delta$, if a node belongs to $\mathcal{I}_{\text{Far}}$, then no matter if the node contains $x'$ or not, it will always stay below $\tau$ i.e., there is no sampled value of $v$ that can help distinguish $X$ and $X'$. The nodes in $\mathcal{I}_{\text{Special}} \cup \mathcal{I}_{\text{Upper-Almost}} \subset \mathcal{I}_{\text{Almost}}$ are the ones that the privacy adversary can take advantage of, and our privacy budget scales linearly with their size. Thus, we do not want the size of $\mathcal{I}_{\text{Special}}$ or $\mathcal{I}_{\text{Upper-Almost}}$ to be large[11]. So we define a good event as

**Definition 3.4.3 (Good Event For A Fixed Output)** Let $c_0$ be as defined in **Algorithm 1**. Fix any full output vector $\vec{b} \in (\perp, \top)^{|\mathcal{H}|}$ and path $(p_1, ..., p_h)$ ordered from leaf to root. Let $\vec{a} := \left(b_{p_1}, ..., b_{p_h}\right) \in (\perp, \top)^h$ be the restriction of $\vec{b}$ to this path. Let $\mathcal{P} := \mathcal{P}(\vec{a})$ be the stopped path from **Definition 3.4.1**. For any vector $\vec{w} = \left(w_p\right)_{p \in \mathcal{P}} \in \mathbb{R}^{|\mathcal{P}|}$, define the good event

$$E(\vec{w}, \vec{a}) := \mathbb{1}[|\mathcal{I}_{\text{Almost}}(\vec{w}, \vec{a})| < c_0]$$

where $\mathcal{I}_{\text{Almost}}(\vec{w}, \vec{a})$ is as defined in **Definition 3.4.2**.

Intuitively, what **Definition 3.4.3** is conveying is that, if we run **Algorithm 1** and get output $\vec{a}$ in the leaf-to-root path containing $x'$ using samples $\vec{v}$ and $\vec{w}$, then the good event is one where $\vec{w}$ and $\vec{a}$ only put a nodes queries in $\mathcal{I}_{\text{Almost}}$. If this event happens with extremely high likelihood, and we can prove privacy under this event, then we can bound the other events to have probability mass at most $\delta$. This is what we do subsequently. We show that that good events are extremely likely by upper bounding the reward of the following game in **Lemma 3.4.4**.

[11] We will be loose in our analysis and upper bound the size $\mathcal{I}_{\text{Almost}}$ instead which may contain a few nodes that are in $\mathcal{I}_{\text{Far}}$ as well. However, we show even $\mathcal{I}_{\text{Almost}}$ is very likely to be small, and this keeps our analysis simpler.

> **Connection with general threshold monitor:** The analysis in Appendix A.2 of [28] requires a more complicated game to accommodate $k > 1$ and arbitrary queries. For us, the order and the queries are fixed, and $k = 1$, so this simpler game (and therefore proof) suffices.

**Lemma 3.4.4 (Coin Flipping Game)** Fix $h \in \mathbb{N}$. Given inputs $\gamma \in \left(0, \frac{1}{2}\right)$ and $\varphi \in \left[\frac{\gamma}{4}, 1-\gamma\right]$ define the following distribution $\mathcal{D}$ over the domain $\{\mathsf{END}, 0, 1\}$ where

$$\Pr_{z \leftarrow\$ \mathcal{D}}[z = 0] = 1 - \gamma - \varphi$$

$$\Pr_{z \leftarrow\$ \mathcal{D}}[z = 1] = \gamma$$

$$\Pr_{z \leftarrow\$ \mathcal{D}}[z = \mathsf{END}] = \varphi$$

We sample a sequence of random variables $Z_1, Z_2, \ldots$ i.i.d. from $\mathcal{D}$ and stop at the random time

$$T := \min(h, \min\{i \geq 1 : Z_i = \mathsf{END}\}),$$

i.e., as soon as we either observe a END or reach $h$ samples, whichever comes first. Let $r := \sum_{i=1}^{T} \mathbb{1}[Z_i = 1]$ be the number of 1's observed before we stop. Then for every integer $c_0 \geq 1$,

$$\Pr[r \geq c_0] \leq \left(\frac{\gamma}{\gamma + \varphi}\right)^{c_0} \leq \left(\frac{4}{5}\right)^{c_0}$$

*Proof.* Fix $c_0 \geq 1$. Set $\rho := \frac{\gamma}{\gamma+\varphi}$, and let $p_{h,c_0}$ be the probability that the game with horizon $h$ yields more than $c_0$ ones. We prove $p_{h,c_0} \leq \rho^{c_0}$ by induction on $h$.

**Base ($h = 0$).** No sample is drawn, so $r = 0$ and $p_{0,c_0} = 0 \leq \rho^{c_0}$.

**Step.** Fix $h \geq 1$ and assume $p_{h-1,c_0} \leq \rho^{c_0}$ for all $c_0 \geq 1$. Condition on the first sample $X_1$; if the game does not stop, the rest is an independent game with horizon $h-1$. With probability $\varphi$ we draw $\perp$ and stop with $r = 0$; with probability $1-\gamma-\varphi$ we draw 0 and still need $c_0$ ones; with probability $\gamma$ we draw 1 and need $c_0 - 1$ more. Hence

$$p_{h,c_0} = (1-\gamma-\varphi)p_{h-1,c_0} + \gamma p_{h-1,c_0-1}. \tag{5}$$

Here $p_{h-1,c_0} \leq \rho^{c_0}$ by hypothesis, and $p_{h-1,c_0-1} \leq \rho^{c_0-1}$ as well (using the boundary $p_{h-1,0} = 1 = \rho^0$ when $c_0 = 1$). With $(\gamma+\varphi)\rho = \gamma$,

$$p_{h,c_0} \leq (1-\gamma-\varphi)\rho^{c_0} + \gamma\rho^{c_0-1} = \rho^{c_0-1}((1-\gamma-\varphi)\rho + \gamma) = \rho^{c_0}. \tag{6}$$

Thus $\Pr[r \geq c_0] \leq \rho^{c_0} = \left(\frac{\gamma}{\gamma+\varphi}\right)^{c_0}$, and $\varphi \geq \frac{\gamma}{4}$ gives $\gamma + \varphi \geq \frac{5\gamma}{4}$, hence $\rho \leq \frac{4}{5}$ and

$$\Pr[r \geq c_0] \leq \left(\frac{4}{5}\right)^{c_0} = e^{-c_0 \ln\left(\frac{5}{4}\right)}. \tag{7}$$

□

Next, we illustrate with **Figure 3** why the game in **Lemma 3.4.4** is relevant for privacy analysis. As we process nodes $(p_1, \ldots, p_{i^\star})$ from leaf to root containing $x'$, the game gives us the random variables $Z_1, \ldots, Z_{i^\star}$. We treat the allocation of nodes to regions in **Figure 3** by $w$ as fixed, and map the action of the random noise $\vec{v}$ to outcomes of the game. If $p_i \in \mathcal{I}_{\text{Special}} \cup \mathcal{I}_{\text{Upper-Almost}}$ and $\vec{v}$ fails to push the noisy count above $\tau$ i.e., $a_{p_i} = \perp$, then $Z_i = 1$. Otherwise, if $a_{p_i} = \perp$ and $p_i \in \mathcal{I}_{\text{Far}}$ then $Z_i = 0$. Finally, we have $Z_{p^\star} = \mathsf{END}$. What the game is really saying is that it is extremely unlikely we will have many nodes in $\mathcal{I}_{\text{Almost}}$ such that $\vec{v}$ fails to push at least one of them over the threshold of $\tau$.

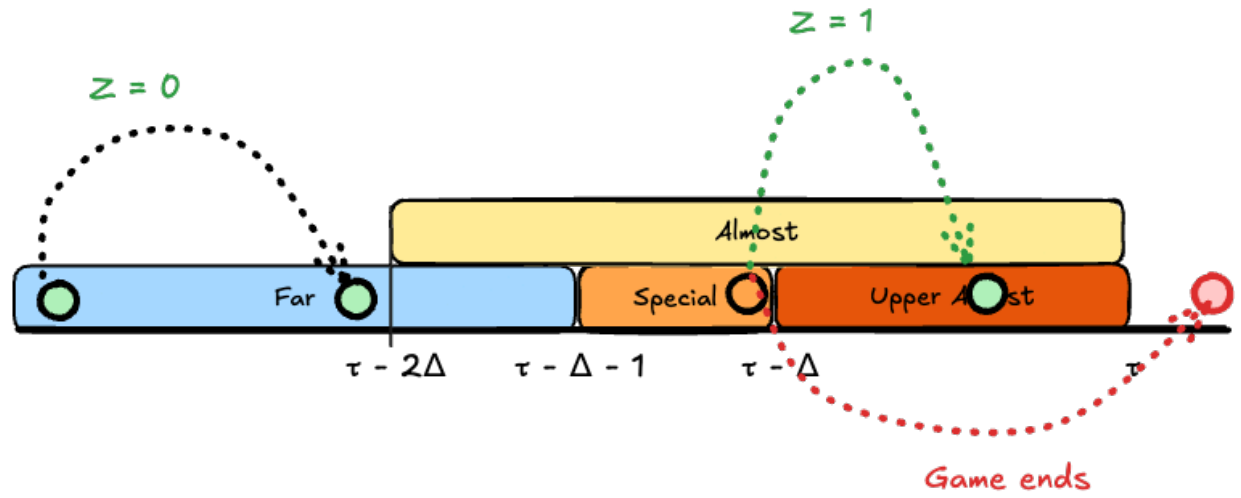

Figure 3: The above figure maps the coin flipping game to the events on $\mathcal{I}_{\text{Active}}$. The crux of the privacy argument simply says that once a node is marked to be in $\mathcal{I}_{\text{Upper-Almost}}$, the chances of it staying a $\perp$ are roughly the same as it becoming a $\top$ once $\vec{v}$ is sampled.

**Corollary 3.4.4.1 (Good Event Is Extremely Likely)** Fix privacy parameters $\varepsilon > 0$ and $\delta \in o(\frac{1}{n})$, and set $c_0 := \log_{\frac{5}{4}} \frac{1}{\delta}$. Fix a dataset $X$, a full output vector $\vec{b} \in (\perp, \top)^{|\mathcal{H}|}$, and a leaf-to-root path $(p_1, \ldots, p_h)$. Let $\vec{a} := \left(b_{p_1}, \ldots, b_{p_h}\right)$ and $\mathcal{P} := \mathcal{P}(\vec{a})$ as defined in **Definition 3.4.1**.

$$\Pr_{\substack{\vec{w} \leftarrow\$ \mathsf{Lap}(6\Delta) \\ \vec{v} \leftarrow\$ \mathsf{Lap}(1/\eta)}} [A(X; \vec{w}, \vec{v})_{\mathcal{P}} = \vec{a}_{\mathcal{P}} \wedge \neg E(\vec{w}, \vec{a})] \leq \delta$$

*Proof.* Fix $p \in \mathcal{I}_{\text{Active}}$. Mapping the variables of **Lemma 3.4.4**, we have $Z = 1$ corresponds to the event where $p \in \mathcal{I}_{\text{Almost}}$. $Z = 0$ corresponds to a node being in $\mathcal{I}_{\text{Far}} \setminus \mathcal{I}_{\text{Almost}}$ and $Z = \mathsf{END}$ corresponds to the event that $p = p_{i*}$. Let $W = \left\{w \in \mathbb{R} : w \geq \tau - 2\Delta - F_{X,\mathcal{S}}(p)\right\}$ and $\mathsf{Lap}(6\Delta)_{|W}$ denote the Laplace distribution re-normalised over the set $W$. Similarly, let $v \leftarrow\$ \mathsf{Lap}(1/\eta)_{|\ v<c}$ denote the conditional distribution obtained by conditioning that the domain of the Laplace distribution is restricted to values less than some $c \in \mathbb{R}$.

$$\gamma := \Pr_{\substack{w \leftarrow\$ \mathsf{Lap}(6\Delta) \\ v \leftarrow\$ \mathsf{Lap}(1/\eta)}} \left[F_{X,\mathcal{S}}(p) + w + \vec{v} < \tau \wedge F_{X,\mathcal{S}}(p) + w \geq \tau - 2\Delta\right] \tag{8.1}$$

$$= \Pr_{\substack{w \leftarrow\$ \mathsf{Lap}(6\Delta) \\ v \leftarrow\$ \mathsf{Lap}(1/\eta)}} \left[\tau - F_{X,\mathcal{S}}(p) - 2\Delta \leq w < \tau - F_{X,\mathcal{S}}(p) - \vec{v}\right] \tag{8.2}$$

$$\leq \Pr_{v \leftarrow\$ \mathsf{Lap}(1/\eta)} [v < -\Delta] + \Pr_{\substack{w \leftarrow\$ \mathsf{Lap}(6\Delta) \\ v \leftarrow\$ \mathsf{Lap}(1/\eta)_{|\ v \geq -\Delta}}} \left[\tau - F_{X,\mathcal{S}}(p) - 2\Delta \leq w < \tau - F_{X,\mathcal{S}}(p) - \vec{v} \mid v \geq -\Delta\right] \tag{8.3}$$

$$\leq \frac{1}{4} + \Pr_{w \leftarrow\$ \mathsf{Lap}(6\Delta)} \left[\tau - F_{X,\mathcal{S}}(p) - 2\Delta \leq w < \tau - F_{X,\mathcal{S}}(p) - v \mid v = -\Delta\right] \tag{8.4}$$

$$\leq \frac{1}{4} + \Pr_{w \leftarrow\$ \mathsf{Lap}(6\Delta)} \left[\tau - F_{X,\mathcal{S}}(p) - 2\Delta \leq w < \tau - F_{X,\mathcal{S}}(p) + \Delta\right] \tag{8.5}$$

$$\leq \frac{1}{4} + \frac{1}{4} = \frac{1}{2} \tag{8.6}$$

In **Equation 8.4**, as we are in the business of upper bounds, we can remove the randomness over $v$ by simply observing that after conditioning over $v \geq -\Delta$, the conditional probability is maximal when we fix $v = -\Delta$ (as this maximizes the possible interval on which sampled $w$ can land). **Equation 8.6** comes from the fact that $w \leftarrow\!\!\$\ \mathsf{Lap}(6\Delta)$ and **Equation 8.5** requires $w$ be in an interval of size $3\Delta$, hence easily upper bounded by $1/4$. The bound $\Pr[v < -\Delta] \leq 1/4$ used above relies on the first quartile of $\mathsf{Lap}(1/\eta)$: it holds whenever $\Delta \geq (\ln 2)/\eta$, i.e. for $\eta$ small enough, which is the case in our regime $\eta = \varepsilon/\log(1/\delta)$ with $\delta$ negligible.

$$\varphi := \Pr_{\substack{w \leftarrow\$\ \mathsf{Lap}(6\Delta) \\ v \leftarrow\$\ \mathsf{Lap}(1/\eta)}} \left[F_{X,\mathcal{S}}(p) + w + \overline{v} \geq \tau \wedge F_{X,\mathcal{S}}(p) + w \geq \tau - 2\Delta\right] \tag{9.1}$$

$$= \Pr_{w \leftarrow\$\ \mathsf{Lap}(6\Delta)} \left[F_{X,\mathcal{S}}(p) + w \geq \tau - 2\Delta\right] \cdot \Pr_{\substack{v \leftarrow\$\ \mathsf{Lap}(\frac{1}{\eta}) \\ w \leftarrow\$ \mathsf{Lap}(6\Delta)_{|W}}} \left[w \geq \tau - \overline{v} - F_{X,\mathcal{S}}(p) \mid w \geq \tau - F_{X,\mathcal{S}}(p) - 2\Delta\right] \tag{9.2}$$

$$\geq \gamma \cdot \Pr_{\substack{v \leftarrow\$\ \mathsf{Lap}(1/\eta) \\ w \leftarrow\$ \mathsf{Lap}(6\Delta)_{|W}}} \left[w \geq \tau - \overline{v} - F_{X,\mathcal{S}}(p) \mid w \geq \tau - F_{X,\mathcal{S}}(p) - 2\Delta\right] \tag{9.3}$$

$$\geq \gamma \cdot \Pr_{\substack{v \leftarrow\$\ \mathsf{Lap}(1/\eta) \\ w \leftarrow\$ \mathsf{Lap}(6\Delta)_{|W}}} \left[w \geq \tau - \overline{v} - F_{X,\mathcal{S}}(p) \mid p \in \mathcal{J}_{\text{Almost}}\right] \tag{9.4}$$

$$\geq \gamma \cdot \Pr_{\mathsf{Lap}(1/\eta)} \left[v \geq -\Delta \mid p \in \mathcal{J}_{\text{Almost}}\right] \cdot \Pr_{\substack{v \leftarrow\$ \mathsf{Lap}(1/\eta)_{|v \geq -\Delta} \\ w \leftarrow\$ \mathsf{Lap}(6\Delta)_{|W}}} \left[w \geq \tau - \overline{v} - F_{X,\mathcal{S}}(p) \mid p \in \mathcal{J}_{\text{Almost}} \wedge v \geq -\Delta\right] \tag{9.5}$$

$$= \gamma \cdot \Pr_{v \leftarrow\$\ \mathsf{Lap}(1/\eta)} \left[v \geq -\Delta\right] \cdot \Pr_{\substack{v \leftarrow\$ \mathsf{Lap}(1/\eta)_{|v \geq -\Delta} \\ w \leftarrow\$ \mathsf{Lap}(6\Delta)_{|W}}} \left[w \geq \tau - \overline{v} - F_{X,\mathcal{S}}(p) \mid p \in \mathcal{J}_{\text{Almost}} \wedge v \geq -\Delta\right] \tag{9.6}$$

$$\geq \gamma \cdot \frac{1}{2} \cdot \Pr_{\substack{v \leftarrow\$ \mathsf{Lap}(1/\eta)_{|v > -\Delta} \\ w \leftarrow\$ \mathsf{Lap}(6\Delta)_{|W}}} \left[w \geq \tau - \overline{v} - F_{X,\mathcal{S}}(p) \mid p \in \mathcal{J}_{\text{Almost}} \wedge v > -\Delta\right] \tag{9.7}$$

$$\geq \gamma \cdot \frac{1}{2} \cdot \Pr_{w \leftarrow\$ \mathsf{Lap}(6\Delta)_{|W}} \left[w \geq \tau + \Delta - F_{X,\mathcal{S}}(p) \mid p \in \mathcal{J}_{\text{Almost}} \wedge v = -\Delta\right] \tag{9.8}$$

$$\geq \gamma \cdot \frac{1}{2} \cdot \frac{1}{2} = \frac{\gamma}{4} \tag{9.9}$$

The last inequality comes from the following analysis. The variable $w$ follows $\mathsf{Lap}(6\Delta)$ *conditioned* on $w \geq \ell$, where $\ell := \tau - 2\Delta - F_{X,\mathcal{S}}(p)$ is the left endpoint of $W$. Once we fix $v = -\Delta$, the target event is $\{w \geq \tau + \Delta - F_{X,\mathcal{S}}(p)\} = \{w \geq \ell + 3\Delta\}$, so the factor equals $\Pr[w \geq \ell + 3\Delta \mid w \geq \ell]$. As a function of $\ell$ this is minimised at $\ell \geq 0$, where it equals exactly $e^{-3\Delta/6\Delta} = e^{-0.5} \approx 0.61 \geq 1/2$; for $\ell < 0$ it is only larger. Hence the factor is at least $1/2$ for every node, irrespective of $F_{X,\mathcal{S}}(p)$.

□

**Connection to General Threshold monitor:** In the general threshold monitor algorithm [28], a good event is more complex. The analysis separately bounds $\mathcal{J}_{\text{Upper-Almost}}$ and $\mathcal{J}_{\text{Almost}}$, as it is not guaranteed that every query has $f_{X'}(x') = 1$ (since that algorithm deals with arbitrary queries). It defines a good event to be when both the above conditions hold, and this makes the constants worse in the general algorithm.

Proving good events are likely is not enough, we also need to show that privacy is preserved under good events. We do this next.

**Lemma 3.4.5 (Privacy Conditioned On Good Events)** Fix privacy parameters $\varepsilon > 0$ and $\delta \in o(\frac{1}{n})$, and set $\xi := 2\eta c_0$. Fix neighbouring datasets $X$ and $X' := X \cup \{x'\}$. Fix any full output vector $\vec{b} \in (\bot, \top)^{|\mathcal{H}|}$ and leaf-to-root path $(p_1, ..., p_h)$ containing $x'$. Let $\vec{a} := (b_{p_1}, ..., b_{p_h})$ and $\mathcal{P} := \mathcal{P}(\vec{a})$.

For every vector $\vec{u} = (u_p)_{p\in\mathcal{P}} \in \mathbb{R}^{|\mathcal{P}|}$ satisfying $E(\vec{u}, \vec{a})$, define $\vec{u'} = (u'_p)_{p\in\mathcal{P}}$ coordinate-wise by

$$u'_p := \begin{cases} u_p - 1 & \text{if } a_p = \top \\ u_p & \text{otherwise} \end{cases}$$

Then

$$\Pr_{\vec{v} \leftarrow_\$ \mathsf{Lap}(1/\eta)}\left[A\left(X'; \vec{u'}, \vec{v}\right)_{\mathcal{P}} = \vec{a}_{\mathcal{P}}\right] \le \Pr_{\vec{v} \leftarrow_\$ \mathsf{Lap}(1/\eta)}[A(X; \vec{u}, \vec{v})_{\mathcal{P}} = \vec{a}_{\mathcal{P}}] \le e^{\xi} \Pr_{\vec{v} \leftarrow_\$ \mathsf{Lap}(1/\eta)}\left[A(X'; \vec{u}, \vec{v})_{\mathcal{P}} = \vec{a}_{\mathcal{P}}\right].$$

*Proof.* As $\vec{u}$ and $\vec{a}$ are fixed, we also get our sets $\mathcal{I}_{\text{Almost}}, \mathcal{I}_{\text{Far}}, \mathcal{I}_{\text{Special}}$ and $\mathcal{I}_{\text{Upper-Almost}}$ as defined in **Definition 3.4.2**. Observe that

$$\begin{aligned}\Pr_{\vec{v} \leftarrow_\$ \mathsf{Lap}(1/\eta)}[A(X; \vec{u}, \vec{v})_{\mathcal{P}} = \vec{a}_{\mathcal{P}}] = &\Pr_{v_{p_{i^\star}} \leftarrow_\$ \mathsf{Lap}(1/\eta)}\left[F_{X,\mathcal{S}}(p_{i^\star}) + u_{i^\star} + \overline{v}_{p_{i^\star}} \ge \tau\right] \\ &\cdot \prod_{p\in\mathcal{I}_{\text{Far}}} \Pr_{v_p \leftarrow_\$ \mathsf{Lap}(1/\eta)}\left[F_{X,\mathcal{S}}(p) + u_p + \overline{v}_p = \bot\right] \\ &\cdot \prod_{p\in\mathcal{I}_{\text{Upper-Almost}}\cup\mathcal{I}_{\text{Special}}} \Pr_{v_p \leftarrow_\$ \mathsf{Lap}(1/\eta)}\left[F_{X,\mathcal{S}}(p) + u_p + \overline{v}_p = \bot\right]\end{aligned} \tag{10.1}$$

The products in **Equation 10.1** comes from the fact that given $\mathcal{P}$ each output along the path is decided independently of the others.

**Upper Bound**: We will bound each product term individually.

*Inflexion Point.* Now as $F_{X',\mathcal{S}'}(p_{i^\star}) > F_{X,\mathcal{S}}(p_{i^\star})$ the following is immediately true

$$\Pr_{v_{p_{i^\star}} \leftarrow_\$ \mathsf{Lap}(1/\eta)}\left[F_{X,\mathcal{S}}(p_{i^\star}) + u_{i^\star} + \overline{v}_{p_{i^\star}} \ge \tau\right] \le \Pr_{v_{p_{i^\star}} \leftarrow_\$ \mathsf{Lap}(1/\eta)}\left[F_{X',\mathcal{S}'}(p_{i^\star}) + u_{i^\star} + \overline{v}_{p_{i^\star}} \ge \tau\right] \tag{11}$$

*Far nodes.* Observe that as $\overline{v}_p \le \Delta$, for all $p \in \mathcal{I}_{\text{Far}}$,

$$\Pr_{v_p \leftarrow_\$ \mathsf{Lap}(1/\eta)}\left[F_{X,\mathcal{S}}(p) + u_p + \overline{v}_p = \bot\right] = \Pr_{v_p \leftarrow_\$ \mathsf{Lap}(1/\eta)}\left[F_{X',\mathcal{S}'}(p) + u_p + \overline{v}_p = \bot\right] = 1 \tag{12}$$

*Upper Almost nodes.* For any $p \in \mathcal{I}_{\text{Upper-Almost}}$, we have $F_{X,\mathcal{S}}(p) + u_p \ge \tau - \Delta$, so for $\Pr_{v_p \leftarrow_\$ \mathsf{Lap}(1/\eta)}\left[F_{X,\mathcal{S}}(p) + u_p + \overline{v}_p = \bot\right] > 0$ we need

$$v_p \le \tau - \left(F_{X,\mathcal{S}}(p) + u_p\right) \le \Delta \tag{13}$$

Using that $\Pr_{v_p \leftarrow_\$ \mathsf{Lap}(1/\eta)}\left[v_p < c\right] \le e^{\eta} \Pr_{v_p \leftarrow_\$ \mathsf{Lap}(1/\eta)}\left[v_p < c - 1\right]$ for any $c$ (shifting the threshold by the sensitivity 1), this gives us

$$\begin{aligned}\Pr_{v_p \leftarrow_\$ \mathsf{Lap}(1/\eta)}\left[F_{X,\mathcal{S}}(p) + u_p + \overline{v}_p = \bot\right] &= \Pr_{v_p \leftarrow_\$ \mathsf{Lap}(1/\eta)}\left[v_p < \tau - \left(F_{X,\mathcal{S}}(p) + u_p\right)\right] &(14.1)\\ &\le e^{\eta} \Pr_{v_p \leftarrow_\$ \mathsf{Lap}(1/\eta)}\left[v_p < \tau - \left(F_{X,\mathcal{S}}(p) + u_p\right) - 1\right] &(14.2)\\ &= e^{\eta} \cdot \Pr_{v_p \leftarrow_\$ \mathsf{Lap}(1/\eta)}\left[F_{X',\mathcal{S}'}(p) + u_p + v_p < \tau\right] &(14.3)\end{aligned}$$

$$\leq e^{\eta} \cdot \Pr_{v_p \leftarrow\$ \mathsf{Lap}(1/\eta)}\left[F_{X',\mathcal{S}'}(p) + u_p + \overline{v}_p < \tau\right] \tag{14.4}$$

$$\leq e^{2\eta} \cdot \Pr_{v_p \leftarrow\$ \mathsf{Lap}(1/\eta)}\left[F_{X',\mathcal{S}'}(p) + u_p + \overline{v}_p < \tau\right] \tag{14.5}$$

*Special nodes.* The special nodes bridge between the far and upper-almost nodes, and so require a slightly different treatment. It is the key part of the proof that depends on the value of $\eta$. For any $p \in \mathcal{I}_{\text{Special}}$, as $\eta := \frac{\varepsilon}{\log(\frac{1}{\delta})} \in (0,1)$ (since $\delta$ is cryptographically negligible and $\varepsilon$ is a small real constant)

$$\Pr_{v_p \leftarrow\$ \mathsf{Lap}(1/\eta)}\left[F_{X,\mathcal{S}}(p) + u_p + \overline{v}_p = \bot\right] \leq 1 \tag{15.1}$$

$$\leq e^{\eta}\left(1 - \frac{\eta}{2}\right) \tag{15.2}$$

$$\leq e^{\eta} \Pr_{v_p \leftarrow\$ \mathsf{Lap}(1/\eta)}\left[v_p < \Delta\right] \tag{15.3}$$

$$\leq e^{\eta} \Pr_{v_p \leftarrow\$ \mathsf{Lap}(1/\eta)}\left[F_{X,\mathcal{S}}(p) + u_p + v_p < \tau\right] \tag{15.4}$$

$$\leq e^{2\eta} \Pr_{v_p \leftarrow\$ \mathsf{Lap}(1/\eta)}\left[F_{X',\mathcal{S}'}(p) + u_p + v_p < \tau\right] \tag{15.5}$$

$$\leq e^{2\eta} \Pr_{v_p \leftarrow\$ \mathsf{Lap}(1/\eta)}\left[F_{X',\mathcal{S}'}(p) + u_p + \overline{v}_p < \tau\right] \tag{15.6}$$

**Equation 15.2** follows from the fact that for any $\eta \in (0,1)$, we have $e^{\eta}(1 - \frac{\eta}{2}) > 1$. **Equation 15.3** comes from the fact that $\Delta := \frac{1}{\eta}\log\left(\frac{1}{\eta}\right)$, and by tail bounds we have $\Pr_{v_i \leftarrow\$ \mathsf{Lap}(1/\eta)}[v_i \geq \Delta] \leq \frac{\eta}{2}$.

We have handled all four types of nodes that appear in **Equation 10.1**. Combining them all

$$\Pr_{\vec{v} \leftarrow\$ \mathsf{Lap}(1/\eta)}[A(X;\vec{u},\vec{v})_{\mathcal{P}} = \vec{a}_{\mathcal{P}}] \leq e^{2\eta\,|\mathcal{I}_{\text{Upper-Almost}} \cup \mathcal{I}_{\text{Special}}|} \Pr_{\vec{v} \leftarrow\$ \mathsf{Lap}(1/\eta)}\left[A(X';\vec{u},\vec{v})_{\mathcal{P}} = \vec{a}_{\mathcal{P}}\right] \tag{16.1}$$

$$\leq e^{2\eta\,|\mathcal{I}_{\text{Almost}}|} \Pr_{\vec{v} \leftarrow\$ \mathsf{Lap}(1/\eta)}\left[A(X';\vec{u},\vec{v})_{\mathcal{P}} = \vec{a}_{\mathcal{P}}\right] \tag{16.2}$$

$$\leq e^{2\eta c_0} \Pr_{\vec{v} \leftarrow\$ \mathsf{Lap}(1/\eta)}\left[A(X';\vec{u},\vec{v})_{\mathcal{P}} = \vec{a}_{\mathcal{P}}\right] \tag{16.3}$$

**Equation 16.3** comes from our assumption that $E(\vec{u},\vec{a}) = 1$, and this completes our proof of the upper bound.

**Lower Bound**: The lower bound is noticeably simpler. For the inflexion node $p^{\star}$ we have,

$$\Pr_{v_{p_{i^\star}} \leftarrow\$ \mathsf{Lap}(1/\eta)}\left[F_{X',\mathcal{S}'}(p_{i^\star}) + u'_{i^\star} + \overline{v}_{p_{i^\star}} \geq \tau\right] = \Pr_{v_{p_{i^\star}} \leftarrow\$ \mathsf{Lap}(1/\eta)}\left[F_{X',\mathcal{S}'}(p_{i^\star}) + u_{i^\star} - 1 + \overline{v}_{p_{i^\star}} \geq \tau\right] \tag{17.1}$$

$$= \Pr_{v_{p_{i^\star}} \leftarrow\$ \mathsf{Lap}(1/\eta)}\left[F_{X,\mathcal{S}}(p_{i^\star}) + u_{i^\star} + \overline{v}_{p_{i^\star}} \geq \tau\right] \tag{17.2}$$

For $p \in \left(\mathcal{I}_{\text{Far}} \cup \mathcal{I}_{\text{Upper-Almost}} \cup \mathcal{I}_{\text{Special}}\right)$, we have (deterministically) $a_p = \bot$ so $u'_p = u_p$, and

$$\Pr_{v_p \leftarrow\$ \mathsf{Lap}(1/\eta)}\left[F_{X',\mathcal{S}'}(p) + u'_p + \overline{v}_p < \tau\right] = \Pr_{v_p \leftarrow\$ \mathsf{Lap}(1/\eta)}\left[F_{X,\mathcal{S}}(p) + 1 + u_p + \overline{v}_p < \tau\right] \tag{18.1}$$

$$\leq \Pr_{v_p \leftarrow\$ \mathsf{Lap}(1/\eta)}\left[F_{X,\mathcal{S}}(p) + u_p + \overline{v}_p < \tau\right] \tag{18.2}$$

Putting all these together, we can conclude,

$$\Pr_{\vec{v} \leftarrow\$ \mathsf{Lap}(1/\eta)}\left[A\left(X';\vec{u'},\vec{v}\right)_{\mathcal{P}} = \vec{a}_{\mathcal{P}}\right] \leq \Pr_{\vec{v} \leftarrow\$ \mathsf{Lap}(1/\eta)}[A(X;\vec{u},\vec{v})_{\mathcal{P}} = \vec{a}_{\mathcal{P}}] \tag{19}$$

This completes the proof.

□

### 3.5 Release On $\mathcal{I}_{\text{Active}}$ is private

We have already shown that if restricted to nodes in $\mathcal{I}_{\text{Active}}$, and conditioning on good events, we have $\xi$ -privacy. Now we remove the conditioning, and prove that the release is $(\xi, \delta)$ private, still restricting to nodes in $\mathcal{I}_{\text{Active}}$. Once this is done, the full privacy over the entire output will immediately follow and is given in **Section 3.6**.

**Lemma 3.5.1 (Privacy Upper Bound)** Fix privacy parameters $\varepsilon > 0$ and $\delta \in o\left(\frac{1}{n}\right)$, and set $\xi := 2\eta c_0$. Fix neighbouring datasets $X$ and $X' := X \cup \{x'\}$. Fix any full output vector $\vec{b} \in (\perp, \top)^{|\mathcal{H}|}$ and leaf-to-root path $(p_1, ..., p_h)$ containing $x'$. Let $\vec{a} := \left(b_{p_1}, ..., b_{p_h}\right)$ and $\mathcal{P} := \mathcal{P}(\vec{a})$.

$$\Pr_{\substack{\vec{w} \leftarrow\$ \,\mathsf{Lap}(6\Delta) \\ \vec{v} \leftarrow\$ \,\mathsf{Lap}(1/\eta)}}[A(X; \vec{w}, \vec{v})_{\mathcal{P}} = \vec{a}_{\mathcal{P}}] \le e^{\xi} \Pr_{\substack{\vec{w} \leftarrow\$ \,\mathsf{Lap}(6\Delta) \\ \vec{v} \leftarrow\$ \,\mathsf{Lap}(1/\eta)}}\left[A(X'; \vec{w}, \vec{v})_{\mathcal{P}} = \vec{a}_{\mathcal{P}}\right] + \delta$$

*Proof.* The noise samples on $\mathcal{P}$ are drawn independently of the data, with $\vec{w} \leftarrow\$ \,\mathsf{Lap}(6\Delta)$ and $\vec{v} \leftarrow \$ \,\mathsf{Lap}(1/\eta)$ i.i.d. per node. For fixed stopped path $\mathcal{P}$ and output $\vec{a}$ we have

$$\begin{aligned}
\Pr_{\substack{\vec{w} \leftarrow\$ \,\mathsf{Lap}(6\Delta) \\ \vec{v} \leftarrow\$ \,\mathsf{Lap}(1/\eta)}}[A(X; \vec{w}, \vec{v})_{\mathcal{P}} = \vec{a}_{\mathcal{P}}] &= \Pr_{\substack{\vec{w} \leftarrow\$ \,\mathsf{Lap}(6\Delta) \\ \vec{v} \leftarrow\$ \,\mathsf{Lap}(1/\eta)}}[A(X; \vec{w}, \vec{v})_{\mathcal{P}} = \vec{a}_{\mathcal{P}} \wedge E(\vec{w}, \vec{a})] && (20.1) \\
&\quad + \Pr_{\substack{\vec{w} \leftarrow\$ \,\mathsf{Lap}(6\Delta) \\ \vec{v} \leftarrow\$ \,\mathsf{Lap}(1/\eta)}}[A(X; \vec{w}, \vec{v})_{\mathcal{P}} = \vec{a}_{\mathcal{P}} \wedge \neg E(\vec{w}, \vec{a})] \\
&\le \Pr_{\substack{\vec{w} \leftarrow\$ \,\mathsf{Lap}(6\Delta) \\ \vec{v} \leftarrow\$ \,\mathsf{Lap}(1/\eta)}}[A(X; \vec{w}, \vec{v})_{\mathcal{P}} = \vec{a}_{\mathcal{P}} \wedge E(\vec{w}, \vec{a})] + \delta && (20.2)
\end{aligned}$$

**Equation 20.2** comes by applying **Corollary 3.4.4.1**. Since $E(\vec{w}, \vec{a})$ is determined by $\vec{w}$ and the fixed output vector $\vec{a}$, we can expand the good-event term by conditioning only on the $w$-samples: $E(\vec{u}, \vec{a})$. Since $\vec{w}$ is continuous, let $f_{\vec{w}}$ denote the density of $\vec{w} \leftarrow\$ \mathsf{Lap}(6\Delta)^{\otimes |\mathcal{P}|}$. Then

$$\begin{aligned}
&\Pr_{\substack{\vec{w} \leftarrow\$ \,\mathsf{Lap}(6\Delta) \\ \vec{v} \leftarrow\$ \,\mathsf{Lap}(1/\eta)}}[A(X; \vec{w}, \vec{v})_{\mathcal{P}} = \vec{a}_{\mathcal{P}} \wedge E(\vec{w}, \vec{a})] && (21.1) \\
&= \int_{\vec{u} \in \mathbb{R}^{|\mathcal{P}|} : E(\vec{u}, \vec{a})} f_{\vec{w}}(\vec{u}) \cdot \Pr_{\vec{v} \leftarrow\$ \,\mathsf{Lap}(1/\eta)}[A(X; \vec{u}, \vec{v})_{\mathcal{P}} = \vec{a}_{\mathcal{P}}] \,\mathrm{d}\vec{u} && (21.2) \\
&\le \int_{\vec{u} \in \mathbb{R}^{|\mathcal{P}|} : E(\vec{u}, \vec{a})} f_{\vec{w}}(\vec{u}) \cdot e^{\xi} \Pr_{\vec{v} \leftarrow\$ \,\mathsf{Lap}(1/\eta)}\left[A(X'; \vec{u}, \vec{v})_{\mathcal{P}} = \vec{a}_{\mathcal{P}}\right] \mathrm{d}\vec{u} && (21.3) \\
&\le e^{\xi} \int_{\vec{u} \in \mathbb{R}^{|\mathcal{P}|}} f_{\vec{w}}(\vec{u}) \cdot \Pr_{\vec{v} \leftarrow\$ \,\mathsf{Lap}(1/\eta)}\left[A(X'; \vec{u}, \vec{v})_{\mathcal{P}} = \vec{a}_{\mathcal{P}}\right] \mathrm{d}\vec{u} && (21.4) \\
&= e^{\xi} \Pr_{\substack{\vec{w} \leftarrow\$ \,\mathsf{Lap}(6\Delta) \\ \vec{v} \leftarrow\$ \,\mathsf{Lap}(1/\eta)}}\left[A(X'; \vec{w}, \vec{v})_{\mathcal{P}} = \vec{a}_{\mathcal{P}}\right]. && (21.5)
\end{aligned}$$

The first step conditions on $\vec{w} = \vec{u}$ (with $\vec{v}$ independent), the second applies **Lemma 3.4.5** on $E(\vec{u}, \vec{a})$, and the third extends the region to all $\vec{u}$. □

**Lemma 3.5.2 (Privacy Lower Bound)** Fix privacy parameters $\varepsilon > 0$ and $\delta \in o(\frac{1}{n})$, and set $\xi := 2\eta c_0$. Fix neighbouring datasets $X$ and $X' := X \cup \{x'\}$. Fix any full output vector $\vec{b} \in (\bot, \top)^{|\mathcal{H}|}$ and leaf-to-root path $(p_1, ..., p_h)$ containing $x'$. Let $\vec{a} := \left(b_{p_1}, ..., b_{p_h}\right)$ and $\mathcal{P} := \mathcal{P}(\vec{a})$.

$$e^{-\xi}\left(\Pr_{\substack{\vec{w} \leftarrow\$ \,\mathsf{Lap}(6\Delta) \\ \vec{v} \leftarrow\$ \,\mathsf{Lap}(1/\eta)}}\left[A(X'; \vec{w}, \vec{v})_{\mathcal{P}} = \vec{a}_{\mathcal{P}}\right] - \delta\right) \leq \Pr_{\substack{\vec{w} \leftarrow\$ \,\mathsf{Lap}(6\Delta) \\ \vec{v} \leftarrow\$ \,\mathsf{Lap}(1/\eta)}}[A(X; \vec{w}, \vec{v})_{\mathcal{P}} = \vec{a}_{\mathcal{P}}]$$

*Proof.* The noise samples on $\mathcal{P}$ are drawn independently of the data, with $\vec{w} \leftarrow\$ \,\mathsf{Lap}(6\Delta)$ and $\vec{v} \leftarrow\$ \,\mathsf{Lap}(1/\eta)$ i.i.d. per node. Let $\vec{u} \in \mathbb{R}^{|\mathcal{P}|}$ be a vector such that $E(\vec{u}, \vec{a})$ holds. Let $\vec{u}' \in \mathbb{R}^{|\mathcal{P}|}$ be as defined in Lemma 3.4.5 where

$$u'_p := \begin{cases} u_p - 1 & \text{if } a_p = \top \\ u_p & \text{otherwise} \end{cases} \tag{22}$$

Let $f_{\vec{w}}$ denote the joint density of $\vec{w} \leftarrow\$ \mathsf{Lap}(6\Delta)^{\otimes |\mathcal{P}|}$ and $f_{w_p}$ the density of each coordinate $w_p \leftarrow\$ \,\mathsf{Lap}(6\Delta)$. Then,

$$f_{\vec{w}}(\vec{u}) = f_{w_{p^\star}}(u_{p^\star}) \prod_{p \in \mathcal{I}_{\text{Far}} \cup \mathcal{I}_{\text{Almost}}} f_{w_p}(u_p) \tag{23.1}$$

$$= f_{w_{p^\star}}(u_{p^\star}) \prod_{p \in \mathcal{I}_{\text{Far}} \cup \mathcal{I}_{\text{Almost}}} f_{w_p}(u'_p) \tag{23.2}$$

$$\geq e^{-\frac{1}{6\Delta}} f_{w_{p^\star}}(u'_{p^\star}) \prod_{p \in \mathcal{I}_{\text{Far}} \cup \mathcal{I}_{\text{Almost}}} f_{w_p}(u'_p) \tag{23.3}$$

$$= e^{-\frac{1}{6\Delta}} f_{\vec{w}}(\vec{u}') \tag{23.4}$$

$$\geq e^{-\xi} f_{\vec{w}}(\vec{u}') \tag{23.5}$$

The last inequality is due to $\xi = 2\eta c_0 = 2\varepsilon/\log(5/4)$ (as $\eta = \varepsilon/\log(1/\delta)$, $c_0 = \log_{\frac{5}{4}}(1/\delta)$). Since $\eta = \varepsilon/\log(1/\delta) \leq \min(\varepsilon, 1/e)$ for $\delta \in o(1/n)$, $1/(6\Delta) = \eta/(6\log(1/\eta)) \leq \eta/6 \leq \varepsilon/6 \leq 2\varepsilon/\log(5/4) = \xi$.

$$\Pr_{\substack{\vec{w} \leftarrow\$ \,\mathsf{Lap}(6\Delta) \\ \vec{v} \leftarrow\$ \,\mathsf{Lap}(1/\eta)}}[A(X; \vec{w}, \vec{v})_{\mathcal{P}} = \vec{a}_{\mathcal{P}}] \geq \Pr_{\substack{\vec{w} \leftarrow\$ \,\mathsf{Lap}(6\Delta) \\ \vec{v} \leftarrow\$ \,\mathsf{Lap}(1/\eta)}}[A(X; \vec{w}, \vec{v})_{\mathcal{P}} = \vec{a}_{\mathcal{P}} \wedge E(\vec{w}, \vec{a})] \tag{24.1}$$

$$= \int_{\vec{u} \in \mathbb{R}^{|\mathcal{P}|} : E(\vec{u}, \vec{a})} f_{\vec{w}}(\vec{u}) \cdot \Pr_{\vec{v} \leftarrow\$ \,\mathsf{Lap}(1/\eta)}[A(X; \vec{u}, \vec{v})_{\mathcal{P}} = \vec{a}_{\mathcal{P}}] \, \mathrm{d}\vec{u} \tag{24.2}$$

$$\geq e^{-\xi} \int_{\vec{u} \in \mathbb{R}^{|\mathcal{P}|} : E(\vec{u}, \vec{a})} f_{\vec{w}}(\vec{u}') \cdot \Pr_{\vec{v} \leftarrow\$ \,\mathsf{Lap}(1/\eta)}\left[A(X'; \vec{u}', \vec{v})_{\mathcal{P}} = \vec{a}_{\mathcal{P}}\right] \mathrm{d}\vec{u} \tag{24.3}$$

$$= e^{-\xi} \int_{\vec{u}' \in \mathbb{R}^{|\mathcal{P}|} : E(\vec{u}', \vec{a})} f_{\vec{w}}(\vec{u}') \cdot \Pr_{\vec{v} \leftarrow\$ \,\mathsf{Lap}(1/\eta)}\left[A(X'; \vec{u}', \vec{v})_{\mathcal{P}} = \vec{a}_{\mathcal{P}}\right] \mathrm{d}\vec{u}' \tag{24.4}$$

$$= e^{-\xi} \Pr_{\substack{\vec{w} \leftarrow\$ \,\mathsf{Lap}(6\Delta) \\ \vec{v} \leftarrow\$ \,\mathsf{Lap}(1/\eta)}}\left[A(X'; \vec{w}, \vec{v})_{\mathcal{P}} = \vec{a}_{\mathcal{P}} \wedge E(\vec{w}, \vec{a})\right] \tag{24.5}$$

$$\geq e^{-\xi} \Pr_{\substack{\vec{w} \leftarrow\$ \,\mathsf{Lap}(6\Delta) \\ \vec{v} \leftarrow\$ \,\mathsf{Lap}(1/\eta)}}\left[A(X'; \vec{w}, \vec{v})_{\mathcal{P}} = \vec{a}_{\mathcal{P}} \wedge E(\vec{w}, \vec{a})\right]$$

$$+e^{-\xi}\left(\Pr_{\substack{\vec{w}\leftarrow\$\,\mathsf{Lap}(6\Delta)\\ \vec{v}\leftarrow\$\,\mathsf{Lap}(1/\eta)}}\left[A(X';\vec{w},\vec{v})_{\mathcal{P}}=\vec{a}_{\mathcal{P}}\wedge\overline{E}(\vec{w},\vec{a})\right]-\delta\right) \tag{24.6}$$

$$\geq e^{-\xi}\left(\Pr_{\substack{\vec{w}\leftarrow\$\,\mathsf{Lap}(6\Delta)\\ \vec{v}\leftarrow\$\,\mathsf{Lap}(1/\eta)}}\left[A(X';\vec{w},\vec{v})_{\mathcal{P}}=\vec{a}_{\mathcal{P}}\right]-\delta\right) \tag{24.7}$$

Equation 24.3 comes from Equation 23.5 and the lower bound in Lemma 3.4.5. Equation 24.4 is justified because $E(\vec{u},\vec{a})=E(\vec{u}',\vec{a})$: since $\vec{u}$ and $\vec{u}'$ differ only at the inflexion node $p^\star$ (never in $\mathcal{I}_{\text{Almost}}$, as $a_{p^\star}=\top$), we have $\mathcal{I}_{\text{Almost}}(\vec{u},\vec{a})=\mathcal{I}_{\text{Almost}}(\vec{u}',\vec{a})$, so the translation $\vec{u}\mapsto\vec{u}'$ maps the region onto itself. Equation 24.6 uses $\Pr_{\substack{\vec{w}\leftarrow\$\,\mathsf{Lap}(6\Delta)\\ \vec{v}\leftarrow\$\,\mathsf{Lap}(1/\eta)}}\left[A(X';\vec{w},\vec{v})_{\mathcal{P}}=\vec{a}_{\mathcal{P}}\wedge\overline{E}(\vec{w},\vec{a})\right]\leq\delta$. Although $E$ is the $X$-based good event, $F_{X',\mathcal{S}'}(p)=F_{X,\mathcal{S}}(p)+1$ on the path, so every node of $\mathcal{I}_{\text{Almost}}$ under $X$ remains in $\mathcal{I}_{\text{Almost}}$ under $X'$; hence the bad event $\overline{E}$ is only more likely under $X'$, and Corollary 3.4.4.1 applied to $X'$ bounds it by $\delta$. □

### 3.6 The Full Proof Of Privacy

Now we have all the tools needed for the full privacy proof.

**Theorem 3.6.1 (Privacy)** **Algorithm 1** is $(2\xi,2\delta)$-DP.

*Proof.*

Fix datasets $X$ and $X':=X\cup\{x'\}$. We can safely assume that $f_X(p)\neq 1$ for all $x'\succeq p$ by the stability histograms argument. In more detail, fix any such node $p$. There can be at most $h$ such isolated nodes. As $F_{X,\mathcal{S}}(p)=0$, the algorithm never considers its release. However, $F_{X',\mathcal{S}'}(p)=1$ so the algorithm considers release. The adversary distinguish $X$ from $X'$, only if $F_{X',\mathcal{S}'}(p)+w_p+\overline{v}_p\geq\tau$, equivalently $w_p+\overline{v}_p\geq\tau-1$. Using $\overline{v}_p\leq\Delta$ after clipping and $w_p\leftarrow\$\,\mathsf{Lap}(6\Delta)$, a Laplace tail bound gives

$$\Pr_{w_p\leftarrow\$\,\mathsf{Lap}(6\Delta)}\left[w_p+\Delta\geq\tau-1\right]\leq\frac{\delta}{h} \tag{25}$$

with room to spare, where the inequality comes from threshold guard of **Algorithm 1** forces $\tau=\Omega(\Delta\log(h/\delta))$. A union bound over the at most $h$ isolated nodes gives us that with probability at most $\delta$ the output is not private.

From here on we can safely assume that $x'$ is never the only contributor the residual frequence of any node $p$. Fix output $\vec{a}:=\left(a_p\right)_{p\in\mathcal{H}}$ where each coordinate is either $\bot$ or $\top$. Let $(p_1,\dots,p_h)$ be the leaf-to-root path of $x'$ ordered from leaf to root, and let $\vec{a}_{x'}:=\left(a_{p_1},\dots,a_{p_h}\right)$ be the restriction of $\vec{a}$ to this path (shown in colour in **Figure 2** in the dotted block). Let $i^\star:=i^\star(\vec{a}_{x'})$ and define the stopped path

$$\mathcal{I}_{\text{Active}}:=\mathcal{P}(\vec{a}_{x'}). \tag{26}$$

Define $M:=|\mathcal{H}|$. Let $(\beta_1,\dots,\beta_M)\leftarrow\$A(X)$ and $(\beta'_1,\dots,\beta'_M)\leftarrow\$A(X')$ be the output of **Algorithm 1** on inputs $X$ and $X'$ respectively with randomness over the samples $(\vec{w},\vec{v})\in\mathbb{R}^M$. Now for this fixed $\vec{b}$, we use the three sets $\mathcal{I}_{\text{Unrelated}},\mathcal{I}_{\text{Active}},\mathcal{I}_{\text{After}}$ as described by **Figure 2**.

Let $\mathcal{R}:=\mathcal{H}\setminus\mathcal{I}_{\text{Unrelated}}=\mathcal{I}_{\text{Active}}\cup\mathcal{I}_{\text{After}}$. By Bayes rule, we have

$$\Pr\left[\vec{\beta}=\vec{a}\right]=\Pr\left[\vec{\beta}_{\mathcal{I}_{\text{Unrelated}}}=\vec{a}_{\mathcal{I}_{\text{Unrelated}}}\right]\cdot\Pr\left[\vec{\beta}_{\mathcal{R}}=\vec{a}_{\mathcal{R}}\mid\vec{\beta}_{\mathcal{I}_{\text{Unrelated}}}=\vec{a}_{\mathcal{I}_{\text{Unrelated}}}\right] \tag{27.1}$$

Focusing on the left hand term of the product in **Equation 27.1**, we need only consider $p \in \mathcal{J}_{\text{Unrelated}}$. As the algorithm proceeds bottom up, we have for any $p \in \mathcal{J}_{\text{Unrelated}}$,

$$\begin{aligned}&\Pr\left[\beta_p = a_p \wedge \vec{\beta}_{\mathsf{Succ}(p)} = \vec{a}_{\mathsf{Succ}(p)}\right] \\ &\quad = \Pr\left[\beta_p = a_p \mid \vec{\beta}_{\mathsf{Succ}(p)} = \vec{a}_{\mathsf{Succ}(p)}\right] \cdot \Pr\left[\vec{\beta}_{\mathsf{Succ}(p)} = \vec{a}_{\mathsf{Succ}(p)}\right]. \end{aligned} \tag{28.1}$$

The corresponding local factor is the same under $X'$:

$$\Pr\left[\beta_p = a_p \mid \vec{\beta}_{\mathsf{Succ}(p)} = \vec{a}_{\mathsf{Succ}(p)}\right] = \Pr\left[\beta'_p = a_p \mid \vec{\beta'}_{\mathsf{Succ}(p)} = \vec{a}_{\mathsf{Succ}(p)}\right]. \tag{29.1}$$

To justify **Equation 29.1**, condition on the event $\vec{\beta}_{\mathsf{Succ}(p)} = \vec{a}_{\mathsf{Succ}(p)}$. This fixes exactly which already-processed successors of $p$ have been selected into $\mathcal{S}$, namely those $q \in \mathsf{Succ}(p)$ with $a_q = \top$. Therefore the threshold test at $p$ is determined by the same fresh noise $(w_p, v_p)$ and by the residual count of $p$ after removing this fixed set of selected successors. As the unconditional count of $p$ is the same under $X$ and $X'$, the probabilities are equal.

For brevity of notation define the two unrelated-output events

$$E^X_{\text{Unrel}} := \left\{\vec{\beta}_{\mathcal{J}_{\text{Unrelated}}} = \vec{a}_{\mathcal{J}_{\text{Unrelated}}}\right\} \tag{30.1}$$

$$E^{X'}_{\text{Unrel}} := \left\{\vec{\beta'}_{\mathcal{J}_{\text{Unrelated}}} = \vec{a}_{\mathcal{J}_{\text{Unrelated}}}\right\}. \tag{30.2}$$

Applying **Equation 29.1** bottom up through the hierarchy over $\mathcal{J}_{\text{Unrelated}}$ gives

$$\Pr\left[E^X_{\text{Unrel}}\right] = \Pr\left[E^{X'}_{\text{Unrel}}\right] \tag{31.1}$$

Thus, to prove the theorem it suffices to show that

$$e^{-\xi}\left(\Pr\left[\vec{\beta'}_{\mathcal{R}} = \vec{a}_{\mathcal{R}} \mid E^{X'}_{\text{Unrel}}\right] - \delta\right) \le \Pr\left[\vec{\beta}_{\mathcal{R}} = \vec{a}_{\mathcal{R}} \mid E^X_{\text{Unrel}}\right] \le e^{\xi} \Pr\left[\vec{\beta'}_{\mathcal{R}} = \vec{a}_{\mathcal{R}} \mid E^{X'}_{\text{Unrel}}\right] + \delta \tag{32.1}$$

Now by Bayes rule and the structure of $\mathcal{H}$, we have

$$\Pr\left[\vec{\beta}_{\mathcal{R}} = \vec{a}_{\mathcal{R}} \mid E^X_{\text{Unrel}}\right] = \Pr\left[\vec{\beta}_{\mathcal{J}_{\text{After}}} = \vec{a}_{\mathcal{J}_{\text{After}}} \mid E^X_{\text{Unrel}} \wedge \vec{\beta}_{\mathcal{J}_{\text{Active}}} = \vec{a}_{\mathcal{J}_{\text{Active}}}\right] \Pr\left[\vec{\beta}_{\mathcal{J}_{\text{Active}}} = \vec{a}_{\mathcal{J}_{\text{Active}}} \mid E^X_{\text{Unrel}}\right] \tag{33}$$

Notice that $\vec{a}$ defines $\mathcal{S}$, and given the unrelated-output event and the decomposition above, we have

$$\Pr\left[\vec{\beta}_{\mathcal{J}_{\text{After}}} = \vec{a}_{\mathcal{J}_{\text{After}}} \mid E^X_{\text{Unrel}} \wedge \vec{\beta}_{\mathcal{J}_{\text{Active}}} = \vec{a}_{\mathcal{J}_{\text{Active}}}\right] = \Pr\left[\vec{\beta'}_{\mathcal{J}_{\text{After}}} = \vec{a}_{\mathcal{J}_{\text{After}}} \mid E^{X'}_{\text{Unrel}} \wedge \vec{\beta'}_{\mathcal{J}_{\text{Active}}} = \vec{a}_{\mathcal{J}_{\text{Active}}}\right] \tag{34}$$

as the contribution of $x'$ has been removed in the residual counts. Thus to prove **Equation 32.1** it suffices to show

$$\begin{aligned} e^{-\xi}\left(\Pr\left[\vec{\beta'}_{\mathcal{J}_{\text{Active}}} = \vec{a}_{\mathcal{J}_{\text{Active}}} \mid E^{X'}_{\text{Unrel}}\right] - \delta\right) &\le \Pr\left[\vec{\beta}_{\mathcal{J}_{\text{Active}}} = \vec{a}_{\mathcal{J}_{\text{Active}}} | E^X_{\text{Unrel}}\right] \\ &\le e^{\xi} \Pr\left[\vec{\beta'}_{\mathcal{J}_{\text{Active}}} = \vec{a}_{\mathcal{J}_{\text{Active}}} | \; E^{X'}_{\text{Unrel}}\right] + \delta \end{aligned} \tag{35.1}$$

For the upper inequality in **Equation 35.1**, first observe that conditioning on $E^X_{\text{Unrel}}$ or $E^{X'}_{\text{Unrel}}$ fixes the same unrelated output $\vec{a}_{\mathcal{J}_{\text{Unrelated}}}$, hence fixes which unrelated successors have already been selected into $\mathcal{S}$. Given this fixed unrelated output, the residuals on $\mathcal{J}_{\text{Active}}$ are fixed functions of the data, and the remaining randomness on $\mathcal{J}_{\text{Active}}$ is independent of the randomness used on $\mathcal{J}_{\text{Unrelated}}$. Thus, conditioning on this event plays *no* further role in the privacy analysis (we can pretend it was never there). For the fixed stopped path $\mathcal{J}_{\text{Active}}$, the fresh noise on $\mathcal{J}_{\text{Active}}$ is drawn independently of the data, with $\vec{w} \leftarrow\!\!\$\; \mathsf{Lap}(6\Delta)$ and $\vec{v} \leftarrow\!\!\$\; \mathsf{Lap}(1/\eta)$ i.i.d. per node. Now apply **Lemma 3.5.1** to the leaf-to-root path $(p_1, ..., p_h)$ with the full output vector equal to the fixed output $\vec{a}$. The stopped path $\mathcal{P}$ in **Lemma 3.5.1** is exactly $\mathcal{I}_{\text{Active}}$, so the upper bound follows. Although **Lemma 3.5.1** uses the left hand side of the equations below, in its theorem statement, we are really proving a fact about the conditional here when we apply the lemma. The reason we don't introduce the notation, is because it plays no role in the privacy analysis, and hence we do not pollute the analysis with unnecessary notation.

$$\Pr_{\substack{\vec{w} \leftarrow\$ \, \mathsf{Lap}(6\Delta) \\ \vec{v} \leftarrow\$ \, \mathsf{Lap}(1/\eta)}} \left[ A(X; \vec{w}, \vec{v})_{\mathcal{I}_{\text{Active}}} = \vec{a}_{\mathcal{I}_{\text{Active}}} \right] := \Pr\left[ \vec{\beta}_{\mathcal{I}_{\text{Active}}} = \vec{a}_{\mathcal{I}_{\text{Active}}} \mid E^X_{\text{Unrel}} \right] \tag{36.1}$$

$$\Pr_{\substack{\vec{w} \leftarrow\$ \, \mathsf{Lap}(6\Delta) \\ \vec{v} \leftarrow\$ \, \mathsf{Lap}(1/\eta)}} \left[ A(X'; \vec{w}, \vec{v})_{\mathcal{I}_{\text{Active}}} = \vec{a}_{\mathcal{I}_{\text{Active}}} \right] := \Pr\left[ \vec{\beta}'_{\mathcal{I}_{\text{Active}}} = \vec{a}_{\mathcal{I}_{\text{Active}}} \mid E^{X'}_{\text{Unrel}} \right] \tag{36.2}$$

The matching lower inequality will follow analogously from **Lemma 3.5.2**. □

**Theorem 3.6.2 (Coverage And Error)** Let $\mathcal{S}$ denote the set of hierarchical heavy hitters selected by **Algorithm 1**, and for any $p \in \mathcal{H}$ let $\tilde{f}_X(p) := \sum_{q \in \mathcal{S} \wedge q \succeq p} \tilde{F}_{X,\mathcal{S}}(q)$ be the estimate of its unconditional frequency reconstructed from the released residual counts, with $c_p := |\{q \in \mathcal{S} : q \succeq p\}|$. Set

$$\alpha = \Omega\left( \frac{\log(\frac{1}{\delta})}{\varepsilon} \cdot \log\left( \frac{\log(\frac{1}{\delta})}{\varepsilon} \right) \cdot \left( \log\left[\frac{1}{\delta}\right] + \log\left[\frac{h}{\beta}\right] \right) \right)$$

Then, with probability $1 - \beta$:

1. **(Coverage)** every $p \in \mathcal{S}$ has $F_{\mathcal{S}}(p) \geq \tau - \alpha$, and every $p \notin \mathcal{S}$ has $F_{\mathcal{S}}(p) \leq \tau + \alpha$.
2. **(Relative error)** every $p \in \mathcal{S}$ has

$$\left| \frac{f_X(p) - \tilde{f}_X(p)}{f_X(p)} \right| \leq \frac{2\alpha}{\tau}$$

and every $p \notin \mathcal{S}$ has

$$\left| \frac{f_X(p) - \tilde{f}_X(p)}{f_X(p)} \right| \leq \frac{1}{c_p + 1} + \frac{2\alpha}{\tau}$$

*Proof.* The analysis for coverage and error follows from the tail bounds of the Laplace distribution and the union bound.

**Coverage:** We give exact answers for any node $p$ with $F_{X,\mathcal{S}}(p) = \tilde{F}_{X,\mathcal{S}}(p) = 0$ (by stability histograms). For each level of the tree we have at most $2n$ Laplace variables: the two samples $w_p \leftarrow\$ \, \mathsf{Lap}(6\Delta)$ and $v_p \leftarrow\$ \, \mathsf{Lap}\left(\frac{1}{\eta}\right)$ drawn per node. Since $\Delta = \frac{1}{\eta}\log\left(\frac{1}{\eta}\right) \geq \frac{1}{\eta}$, we have $\frac{1}{\eta} \leq 6\Delta$, so the tail of $v_p$ is dominated by that of $w_p$, namely $\Pr_{v_p \leftarrow\$ \, \mathsf{Lap}\left(\frac{1}{\eta}\right)}\left[|v_p| > t\right] = e^{-\eta t} \leq e^{-\frac{t}{6\Delta}}$; we may therefore treat both as $\mathsf{Lap}(6\Delta)$ samples. Clipping only decreases magnitude, so $|\overline{v}_p| \leq |v_p|$. Taking the union bound over all $2n$ samples at a level with total failure probability $\frac{\beta}{h}$, with probability $1 - \frac{\beta}{h}$ every node satisfies $|w_p|, |\overline{v}_p| \leq 6\Delta \log\left(\frac{2nh}{\beta}\right)$, hence

$$|w_p + \overline{v}_p| \leq |w_p| + |\overline{v}_p| \leq 12\Delta \log\left(\frac{2nh}{\beta}\right) = O\left(\Delta \log\left(\frac{nh}{\beta}\right)\right) \tag{37}$$

Since $\delta = \mathsf{negl}(n)$ gives $n \leq \frac{1}{\delta}$, this is at most $O\left(\Delta \log\left(\frac{h}{\delta\beta}\right)\right) = O\left(\Delta\left(\log\left[\frac{1}{\delta}\right] + \log\left[\frac{h}{\beta}\right]\right)\right)$. So if we set

$$\alpha = \Omega\left(\Delta\left(\log\left[\frac{1}{\delta}\right] + \log\left[\frac{h}{\beta}\right]\right)\right) \tag{38.1}$$

$$= \Omega\left(\frac{1}{\eta}\log\left(\frac{1}{\eta}\right)\left(\log\left[\frac{1}{\delta}\right] + \log\left[\frac{h}{\beta}\right]\right)\right) \tag{38.2}$$

$$= \Omega\left(\frac{\log(\frac{1}{\delta})}{\varepsilon}\cdot\log\left(\frac{\log(\frac{1}{\delta})}{\varepsilon}\right)\cdot\left(\log\left[\frac{1}{\delta}\right] + \log\left[\frac{h}{\beta}\right]\right)\right), \tag{38.3}$$

then taking the union bound over the $h$ levels of the tree, with probability $1-\beta$ the per-node DP noise satisfies $|w_p + \overline{v}_p| \le \alpha$ for every node. Recall that $p \in \mathcal{S}$ if and only if $F_{X,\mathcal{S}}(p) + w_p + \overline{v}_p \ge \tau$, and condition on the good event $|w_p + \overline{v}_p| \le \alpha$ (probability $\ge 1-\beta$). If $p \in \mathcal{S}$, then $F_{X,\mathcal{S}}(p) \ge \tau - (w_p + \overline{v}_p) \ge \tau - \alpha$, so every selected prefix has residual at least $\tau - \alpha$ (no false positives). Conversely, if $p \notin \mathcal{S}$, then $F_{X,\mathcal{S}}(p) < \tau - (w_p + \overline{v}_p) \le \tau + \alpha$. Equivalently, every prefix that is $\alpha$-far from $\tau$ is classified correctly: $F_{X,\mathcal{S}}(p) \ge \tau + \alpha$ forces $p \in \mathcal{S}$, and $F_{X,\mathcal{S}}(p) \le \tau - \alpha$ forces $p \notin \mathcal{S}$.

**Relative Error**

Fix $p \in \mathcal{S}$. We have

$$\sum_{q \in \mathcal{S} \wedge q \succeq p} F_{X,\mathcal{S}}(q) = f_X(p), \tag{39}$$

Therefore,

$$|f_X(p) - \tilde{f}_X(p)| = \left|f_X(p) - \sum_{q \in \mathcal{S} \wedge q \succeq p} \tilde{F}_{X,\mathcal{S}}(q)\right| \tag{40.1}$$

$$\le \underbrace{\left|f_X(p) - \sum_{q \in \mathcal{S} \wedge q \succeq p} F_{X,\mathcal{S}}(q)\right|}_{=0} + \sum_{q \in \mathcal{S} \wedge q \succeq p} |F_{X,\mathcal{S}}(q) - \tilde{F}_{X,\mathcal{S}}(q)| \tag{40.2}$$

$$\le c_p \cdot \alpha = c_p \alpha \tag{40.3}$$

where the last bound uses $|F_{X,\mathcal{S}}(q) - \tilde{F}_{X,\mathcal{S}}(q)| \le \alpha$ on the good event from the coverage analysis, summed over the $c_p$ selected prefixes.

To turn this into a relative error, bound $f_X(p)$ from below. By coverage, each of the $c_p$ selected $q$ with $q \succeq p$ has $F_{X,\mathcal{S}}(q) \ge \tau - \alpha$, and these residual masses are disjoint and all lie under $p$, so

$$f_X(p) = \sum_{q \in \mathcal{S} \wedge q \succeq p} F_{X,\mathcal{S}}(q) \ge c_p(\tau - \alpha). \tag{41}$$

Therefore

$$\left|\frac{f_X(p) - \tilde{f}_X(p)}{f_X(p)}\right| \le \frac{c_p \alpha}{c_p(\tau - \alpha)} = \frac{\alpha}{\tau - \alpha}. \tag{42}$$

Finally, the explicit threshold guard of **Algorithm 1** forces any non-trivial run to satisfy $\tau \ge 24\Delta\log(2h/(\delta\beta))$. The coverage bound gives $\alpha \le 12\Delta\log(2nh/\beta) \le 12\Delta\log(2h/(\delta\beta))$ (using $n \le 1/\delta$), so this guard is exactly the condition $\tau \ge 2\alpha$, i.e. $\alpha \le \tau/2$. Hence $\tau - \alpha \ge \tau/2$, so

$$\frac{\alpha}{\tau - \alpha} \le \frac{2\alpha}{\tau}, \tag{43}$$

Now $p \notin \mathcal{S}$, so $p$ is no longer a selected ancestor of the mass beneath it.

$$f_X(p) - \sum_{q \in \mathcal{S} \wedge q \succeq p} F_{X,\mathcal{S}}(q) = F_{X,\mathcal{S}}(p). \tag{44}$$

The key point is that this miss is small as $F_{X,\mathcal{S}}(p) + w_p + \overline{v}_p < \tau$, so on the good event ($w_p + \overline{v}_p \geq -\alpha$)

$$F_{X,\mathcal{S}}(p) < \tau + \alpha, \tag{45}$$

This gives us,

$$\left|f_X(p) - \tilde{f}_X(p)\right| \leq \underbrace{\left|f_X(p) - \sum_{q \in \mathcal{S} \wedge q \succeq p} F_{X,\mathcal{S}}(q)\right|}_{=F_{X,\mathcal{S}}(p)} + \sum_{q \in \mathcal{S} \wedge q \succeq p} \left|F_{X,\mathcal{S}}(q) - \tilde{F}_{X,\mathcal{S}}(q)\right| \tag{46.1}$$

$$\leq F_{X,\mathcal{S}}(p) + c_p \alpha \tag{46.2}$$

By the residual decomposition,

$$f_X(p) = F_{X,\mathcal{S}}(p) + \sum_{q \in \mathcal{S} \wedge q \succeq p} F_{X,\mathcal{S}}(q) \geq F_{X,\mathcal{S}}(p) + c_p(\tau - \alpha) \tag{47}$$

since the $c_p$ selected residuals under $p$ are disjoint and each at least $\tau - \alpha$. Hence

$$\left|\frac{f_X(p) - \tilde{f}_X(p)}{f_X(p)}\right| \leq \frac{F_{X,\mathcal{S}}(p) + c_p \alpha}{F_{X,\mathcal{S}}(p) + c_p(\tau - \alpha)} \tag{48.1}$$

$$\leq \frac{(\tau + \alpha) + c_p \alpha}{(\tau + \alpha) + c_p(\tau - \alpha)} \tag{48.2}$$

$$\leq \frac{1}{c_p + 1} + \frac{2\alpha}{\tau}, \tag{48.3}$$

where the second line maximises over $F_{X,\mathcal{S}}(p) < \tau + \alpha$ (the ratio increases in $F_{X,\mathcal{S}}(p)$ since $\tau - \alpha > \alpha$) and the third uses $\alpha \leq \tau/2$ from the threshold guard. □

## 4 Streaming Private Hierarchical Heavy Hitters

In the previous section we proved that the scale of the DP noise per query scales only by a small constant factor that is independent of the height of the hierarchy and the number of hierarchical heavy hitters in the dataset. A critical factor to being able to bypass composition was that there was no space limitation, and we could store the *exact* conditional count for *every* query in memory. This meant that we could eventually remove the contribution of the neighbouring element $x'$ completely (restricting the privacy loss entirely to the green and orange nodes in **Figure 2**). Thus for a large majority of queries, neighboring inputs $X$ and $X'$ were treated identically.

**Algorithm 2:** Insert into MG Sketch

**Input:** Next data $x \in \mathcal{H}$, increment $v > 0$. **Parameters:** Number of counters $\kappa$.

1 **if** $x \in \mathcal{T}$ **then**
2 $\quad C[x] = C[x] + v$
3 **else if** $C[i] \geq v$ for all $i \in \mathcal{T}$ **then**
4 $\quad C[i] = C[i] - v$ for all $i \in \mathcal{T}$
5 **else**
6 $\quad$ Let $\tilde{y} = \arg\min_{y \in \mathcal{T}} C[y]$
7 $\quad \mathcal{T} = (\mathcal{T} \setminus \{\tilde{y}\}) \cup \{x\}$
8 $\quad C[x] = v$
9 **end if**

Algorithm 2: Insertion Operation For A Single MG Sketch

In the streaming setting, we do not have the luxury of storing exact counts. Thus, we will need to use some streaming data structure [8, 13, 36] in order to approximate residual frequencies. In this work, we use one Misra Gries (MG) sketch with $\kappa$ counters per level of the hierarchy. Thus the total space used is $O(\kappa h)$. We choose the MG sketch for two reasons. Firstly, Pankaj K Agarwal, Graham Cormode, Zengfeng Huang, Jeff M Phillips, Zhewei Wei, and Ke Yi. [1] showed that the MG sketch is isomorphic to the Space Saving algorithm, and Michael Mitzenmacher, Thomas Steinke, and Justin Thaler. [36] show that the space saving algorithm is optimal for non-private hierarchical heavy hitter estimation. Thus, if we replace SS with MG in the non private HHH estimation problem, we retain the same optimality results. Secondly, with the MG sketch we leverage the structure in the output to circumvent the composition bounds due to approximation. The main bottleneck in approximation algorithms is that the approximated function might have large global sensitivity. Indeed T -H Hubert Chan, Mingfei Li, Elaine Shi, and Wenchang Xu. [8] show that the MG sketch has global sensitivity $\Delta_G = \kappa$. This implies that if we naively used noise scaled by the sensitivity of the function, the DP error grows as the streaming error drops. However we are able to leverage the critical observation made by [30:Lemma 5], who show that if the global sensitivity of the MG sketch is high, then counts of each counter after processing neighbouring streams are highly correlated. Just like in **Section 3** where we made use of monotonicity of residual queries, we will use this correlation to circumvent composition. Our techniques are related to the more general observation that structure has often been used to bypass composition in the privacy community [19, 21, 25, 28]. In summary, to construct our HHH estimation algorithm in the streaming setting, we first modify the Michael Mitzenmacher, Thomas Steinke, and Justin Thaler. [36] HHH algorithm to use the modified MG sketch from Christian Janos Lebeda and Jakub Tetek. [30] instead of the SpaceSaving algorithm. Then we repeatedly leverage the structure of the output to bound the DP error to be independent of the available space. Note, we cannot avoid the approximation error introduced due to lack of space, regardless of privacy. In the streaming setting, our contribution is to show that the DP error for HHH is not affected by this approximation parameter $\kappa$. Before describing our algorithm, we first provide an overview of the proof behind how we keep the DP error independent of the number of counters, and why the tricks used in the non-streaming section to make the noise independent of the height of the hierarchy *no longer apply*.

## 4.1 Technical Overview

Our main insight in the non-streaming setting was that the DP noise could be independent of the height of the hierarchy $h$. Unfortunately, in the streaming setting, this claim no longer holds true. To fully describe why, we first re-state a lemma by [30:Lemma 5] which formally describes the observed outcomes when a MG sketch processes neighbouring input streams.

**Lemma 4.1.1 (Lebeda & Tětek; Lemma 5, restated)** Let $X = X' \cup \{x\}$. Let $(\mathcal{T}, C) \leftarrow \mathsf{MG}(\kappa, X)$ and $(\mathcal{T}', C') \leftarrow \mathsf{MG}(\kappa, X')$ be the output of **Algorithm 2** with inputs $X$ and $X'$. Then, $|\mathcal{T} \cup \mathcal{T}'| \geq \kappa - 2$; for all $x \notin \mathcal{T} \cup \mathcal{T}'$, $C[x] \leq 1$ and $C'[x] \leq 1$; and exactly one of the following is true:

1. $\exists i \in \mathcal{T}$, such that $C[i] = C'[i] + 1$, and $\forall j \neq i : C[j] = C'[j]$.
2. $\forall i \in \mathcal{T}' : C[i] = C'[i] - 1$, and $C'[j] = 0$ for $j \notin \mathcal{T}'$.

In the lemma above, $(\mathcal{T}, C)$ and $(\mathcal{T}', C')$ denote the output of the MG algorithm (**Algorithm 2**) on the two neighbouring streams $X$ and $X'$. The lemma states that there can be at most 2 elements that are in $\mathcal{T}$, but not in $\mathcal{T}'$, and vice versa. We will refer to these as *isolated elements*. Furthermore, the count of an isolated element must always be at most 1. We can use the thresholding trick from stability histograms to suppress isolated elements with high probability. The threshold for suppression is set a little higher, as there are two isolated elements instead of one. Along with the above statements, after processing neighbouring data streams, **Lemma 4.1.1** also states that the resulting sketches could be in one of two scenarios illustrated in **Figure 4**. In the scenario shown in **Figure 4** (b), we have two histograms with $\kappa$ bins that differ in at most two locations by a count of 1. All other bins have the exact same values across both sketches. We can easily release private versions of these histograms by simply applying the Laplace mechanism with suppression of small values. The other scenario, described by **Figure 4** (a), appears more problematic at first glance. We have that the counts across two outputs $C$ and $C'$ in the two sketches all differ by the *same* amount across *all* the bins. This is undesirable in two ways. First, now the global sensitivity of the approximated counts is $\kappa$, so it appears that the DP noise will need to scale linearly in $\kappa$ (which can be a large constant). Secondly, notice that we can no longer restrict the influence of differing nodes in $\mathcal{T}'$ to a single hierarchical heavy hitter like we did in the illustration given in **Figure 2**. The neighbouring elements are now spread across all the bins, and they influence *multiple* hierarchical heavy hitters. Furthermore, as the MG sketch often under-approximates the true count of an element, we cannot guarantee that we have removed the influence of a neighbouring element higher up in the tree by removing a descendant lower in the tree. This means we cannot restrict the influence of neighbouring elements and partition the tree like before to avoid paying the privacy loss due to composition for $h$ levels of the hierarchy. Therefore, it seems challenging to circumvent composition bounds due to the height when using the MG sketch.

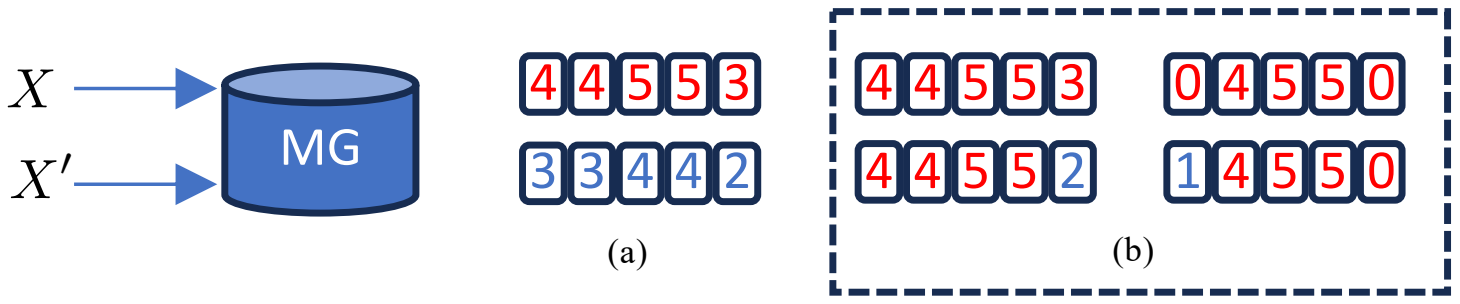

Figure 4: The figure above illustrates the implications of **Lemma 4.1.1**. There are two possible configurations after processing neighbouring streams. One configuration, depicted by figure (a), is that every count in one sketch is different from the count in the other sketch by one unit in the same direction. The other outcome is that either exactly one counter is different for all counters that are non-zero in both sketches. Additionally, when in configurations depicted by Figure (b), there can be at most 2 elements per sketch that are not in the other sketch. The count of these elements when present is always 1. The second outcome, denoted by figure (b) is identical to the configuration discussed for stability histograms in the previous section, where each sketch has at most a single count that is different.

Although we cannot remove the dependence on the height of the hierarchy in the streaming setting, inspired by the observations made by Christian Janos Lebeda and Jakub Tetek. [30], we are still able to remove the dependence on $\kappa$. We show that, even in the scenario depicted by **Figure 4** (a), the DP noise is actually independent of the number of counters in the sketch. This might appear unintuitive at first glance, as the global sensitivity is still $\kappa$. To gain intuition towards understanding why, we first review the privacy proof of the original SVT algorithm. The main observation there was that before we see the first $\top$, we could group all the small valued queries for which the output was $\bot$, and treat them as *one* query, thus paying for them only *once*. We could do so because the event that all of those queries being small is equivalent to saying that the maximum of all the those queries is small. Given two neighbouring datasets, we might have $t-1$ queries, each with sensitivity 1, and thus a total sensitivity of $t-1$, but the max of these queries is still a single query with sensitivity 1. Thus, we have a *single* equivalent query that captures the event described by $t-1$ queries with output $\bot$. A similar reframing also applies to the sketches produced by neighbouring streams. Observe that despite $\kappa$ bins being different in $C$ and $C'$, the direction in which they are different and the amount by which they are different is the same for *all* bins. Thus, we have a guarantee that, if one of the bins in $C$ and $C'$ is different, *all* bins are different, and importantly in the exact *same* way. There is actually just one degree of freedom, despite there being $\kappa$ different counts to consider. Once one bin differs by one, *all* bins differ by one (i.e. the counts are correlated). In the Above threshold (a single invocation of the SVT algorithm), we looked at the max query, and in this case we can look at any one query (say the first one, which reveals everything about the other ones). The effect is equivalent. In the proof for **Theorem 4.1.2** we formally show how we can handle all $\kappa$ counts being different with just one sample of noise, as if we had just a single query.

> **Remark:** [30:Lemma 6] make the *same* claim to use one sample to cover all bins. Although the final claim is correct, there is a minor issue in their proof. In their proof, the authors define the function $g : \mathbb{R} \to \mathbb{R}^k$ such that $g(a) = a1^k$. Thus, $g^{-1}(\vec{x})$ is defined only if *all* coordinates of $\vec{x}$ are the *same*. **Lemma 4.1.1** does not guarantee that all the counts are the same, only that two neighbouring differ by the same amount in the same direction. However, their proof [30] relies on inverting $g^{-1}(\vec{x})$ for general $x \in \mathbb{R}^k$, which is undefined. In our analysis, there is no need to define such a function $g$ and we can obtain our results using the above observations.

**Algorithm 4** describes our algorithm for computing hierarchical private heavy hitters. First, we compute noisy heavy hitters at each level of the hierarchy using **Algorithm 3**. We show that the output of **Algorithm 3** is private. Then we conservatively post-process this private output to get our desired result.

> **Remark:** For a fixed level $\ell$, **Algorithm 3** is *identical* to [30:Algorithm 2], with an essential alteration. In the count-release step of the algorithm, we generate a fresh batch of randomness independent of thresholding operation to release approximate counts. The reason for this is subtle but immediate when viewing the algorithm from the lens of the SVT (where this issue is well documented). The construction described by Christian Janos Lebeda and Jakub Tetek. [30] re-uses the thresholding noise and it is therefore *not* $(\varepsilon, \delta)$ differentially private. The intuition for this is the following: If we re-use the noise in the thresholding/release lines, then we reveal partial information about the global sample $\gamma_l$ *every* time we release a noisy count. With every release, we restrict the possible values $\gamma_l$ could take, thereby shrinking the variance of the privacy distribution. We pay for this shrinkage with a reduced privacy budget, in that the algorithm is now $\varepsilon'$ private for $\varepsilon' > \varepsilon$. See **Appendix D** for more details about how our privacy proof would break if we re-used noise.

**Algorithm 3:** Private Release

**Input:** Data stream $X$, Privacy parameters $\varepsilon$ and $\delta$.

1. Construct $h$ sketches $(\mathsf{MG}_1, ..., \mathsf{MG}_h)$ by running **Algorithm 2** for each $x \in X$ and every generalization of $x$.
2. **for** $\ell \in [h, h-1, ..., 1]$ **do**
3. $\gamma_\ell \leftarrow \mathsf{Laplace}(\frac{2h}{\varepsilon})$
4. Let $(\mathcal{T}_\ell, C_\ell) = \mathsf{MG}_\ell$
5. **for** $i \in \mathcal{T}_\ell$ **do**
6. $w_i \leftarrow \mathsf{Laplace}(\frac{4h}{\varepsilon})$
7. **if** $C_\ell[i] + \gamma_\ell + w_i > 1 + \frac{6h}{\varepsilon}\log(3h/\delta)$ **then**
8. $C_\ell[i] = C_\ell[i] + \mathsf{Laplace}(\frac{4h}{\varepsilon})$ (fresh noise)
9. **else**
10. $C_\ell[i] = 0$
11. **end if**
12. **end for**
13. Set $\overline{\mathsf{MG}_\ell} = (\mathcal{T}_\ell, C_\ell)$
14. **end for**
15. Output $\left(\overline{\mathsf{MG}_1}, ..., \overline{\mathsf{MG}_h}\right)$.

Algorithm 3: Private Release

**Theorem 4.1.2 (Privacy)** **Algorithm 3** is $(\varepsilon, \delta)$-DP.

*Proof.* Fix some level $\ell$. At each level of the tree, we have at most $\kappa + 1$ independent samples from a Laplace distribution, denoted by $w_1, ..., w_\kappa$ and $\gamma_\ell$ in **Algorithm 3**. We also process each level of the data structure independently. The resulting output $(\mathcal{T}_\ell, C_\ell)$ or $(\mathcal{T}_{\ell'}, C_{\ell'})$ is in one of two conditions stated in **Lemma 4.1.1** and as illustrated in **Figure 4**. We will use the first $\kappa$ independent samples to handle the case in **Figure 4** (b), which is essentially the stability histogram problem with Laplace noise and global sensitivity 2.

**Case 1.** First we need to handle the isolated counters in stability histograms. Remember that in isolated counter events there is an element $x \in \mathcal{T}_\ell$ but $x \notin \mathcal{T}_{\ell'}$, or vice versa. From **Lemma 4.1.1**, as $|\mathcal{T} \cap \mathcal{T}'| \geq \kappa - 2$, there can be at most 2 isolated elements included in the output of the algorithm at any given level $\ell$. Let $a, b \in \mathcal{T}_\ell$ denote the possible isolated elements at level $\ell$ after processing $X$. If $a$ or $b$ were to ever show up in the output, the privacy adversary could perfectly distinguish $X$ and $X'$. For convenience, define $\upsilon := \frac{2h}{\varepsilon}\log(\frac{3h}{\delta})$. For $a$ to show up in the final output, we need $w_a + \gamma_\ell$ to push the noisy count of $a$ over $1 + 3\upsilon$. Once again from the tail bounds of the Laplace distribution, we have that the probability $\Pr[w_a > 2\upsilon] \leq \frac{\delta}{3h}$. A similar argument bounds the probability of either of $\gamma_\ell$ and $w_b$ exceeding $\upsilon$ and $2\upsilon$, respectively. Using the same union bound across both events, we get with probability at least $1 - \delta/h$ that both $w_a + \gamma_\ell \leq 3\upsilon$ and $w_b + \gamma_\ell \leq 3\upsilon$. For all counters, we add noise drawn from a Laplace distribution, using privacy parameter $\varepsilon/4h$ once during thresholding, and once during public release. Taking the union over the height of the hierarchy, we get that the first case is handled with $(\varepsilon, \delta)$-privacy.

**Case 2.** Next we show that the other case (**Figure 4** (a)), where all counters in $C$ and $C'$ are off by the *same* amount in the *same* direction, is also handled with $\varepsilon/2h$-DP. This is case (2) of **Lemma 4.1.1**, where $C[i] = C'[i] - 1$ for all $i \in \mathcal{T}'$. Remember that our analysis considered the noise in $w_1, ..., w_\kappa$ already for stability histograms. Thus we want $\gamma_\ell$ to cover for this other case entirely. Here we use notation $a1^\kappa$ to denote the vector $[a, a, ..., a]$ of size $\kappa$ for any $a \in \mathbb{R}$, and $\vec{w}$ succinctly denotes all the $w_i$'s at level $\ell$. The analysis below is for any fixed $\ell \in [h]$, so we drop the subscript $\ell$ to avoid an overload of notation. Let $\vec{y} \in \{\perp, \top\}^\kappa$ be the released-or-not pattern, with coordinate $i$ equal to $\top$ exactly when the noisy count exceeds the threshold $1 + \frac{6h}{\varepsilon}\log\left(\frac{3h}{\delta}\right)$. Since $C = C' - 1^\kappa$,

$$\Pr_{\substack{\gamma \leftarrow\$ \mathsf{Lap}((2h)/\varepsilon) \\ \vec{w} \leftarrow\$ \mathsf{Lap}((4h)/\varepsilon)^{\otimes\kappa}}} [C + \gamma 1^\kappa + \vec{w} = \vec{y}] \tag{49.1}$$

$$= \int f_{\vec{w}}(\vec{z}) \cdot \Pr_{\gamma \leftarrow\$ \mathsf{Lap}((2h)/\varepsilon)} [C + \gamma 1^\kappa + z = \vec{y}] \, \mathrm{d}\vec{z} \tag{49.2}$$

$$= \int f_{\vec{w}}(\vec{z}) \cdot \Pr_{\gamma \leftarrow\$ \mathsf{Lap}((2h)/\varepsilon)} [C' + (\gamma - 1) 1^\kappa + z = \vec{y}] \, \mathrm{d}\vec{z} \tag{49.3}$$

$$\leq e^{\varepsilon/2h} \int f_{\vec{w}}(\vec{z}) \cdot \Pr_{\gamma \leftarrow\$ \mathsf{Lap}((2h)/\varepsilon)} [C' + \gamma 1^\kappa + z = \vec{y}] \, \mathrm{d}\vec{z} \tag{49.4}$$

$$= e^{\varepsilon/2h} \Pr_{\substack{\gamma \leftarrow\$ \mathsf{Lap}((2h)/\varepsilon) \\ \vec{w} \leftarrow\$ \mathsf{Lap}((4h)/\varepsilon)^{\otimes\kappa}}} [C' + \gamma 1^\kappa + \vec{w} = \vec{y}] \tag{49.5}$$

$f_{\vec{w}}$ denotes the density of $\mathsf{Lap}\left(4\frac{h}{\varepsilon}\right)^{\otimes\kappa}$. Finally, applying basic composition over $h$ levels of the tree, we get that the entire algorithm restricted to the scenario depicted in **Figure 4** (a) is $(\varepsilon/2, 0)$-DP. Once we have decided which counters will be used, the final noisy release is $(\varepsilon/2, 0)$-DP. Hence, regardless of which case we happen to be in, we obtain that the whole protocol is $(\varepsilon, \delta)$-DP. □

**Algorithm 4:** Final HHH-Streaming Algorithm

**Input:** Data $X$, Privacy parameter $\varepsilon \in (0, \log n)$, $\delta = o(1/n)$, Threshold $\tau > 0$, Confidence Parameter $\beta \in (0, 1/2)$. **Parameters:** Number of counters $\kappa$, Height of hierarchy $h$.

1. Run **Algorithm 3** with input $X$, Threshold $\tau$, and privacy parameters $(\varepsilon, \delta)$ to get outputs $\left(\overline{\mathsf{MG}_1}, ..., \overline{\mathsf{MG}_h}\right)$.
2. $\alpha_1 = \left(1 + \frac{6h \log[3h/\delta]}{\varepsilon}\right) + \frac{n}{\kappa+1} + \left(\frac{8h}{\varepsilon} \log\left[\frac{2\kappa h}{\beta}\right]\right)$
3. $\alpha_2 = \left(1 + \frac{6h \log[3h/\delta]}{\varepsilon}\right) + \left(\frac{8h}{\varepsilon} \log\left[\frac{2\kappa h}{\beta}\right]\right)$
4. **for** $l \in [h, h-1, ..., 1]$ **do**
5.     Let $(\mathcal{T}_l, C_l) = \overline{\mathsf{MG}_l}$
6.     **for** $e \in \mathcal{T}_l$ **do**
7.         **if** $C_l[e] + \alpha_1 > \tau - \alpha_1$ **then**
8.             $\mathcal{S} = \mathcal{S} \cup \{e\}$
9.             $\tilde{f}_X(e) = C_l[e]$    (Noisy Estimate)
10.             **for** $p \in \mathsf{Generalise}(e)$ **do**
11.                 Let $i = \mathsf{Level}(p)$, $(\mathcal{T}_i, C_i) = \mathsf{MG}_i$
12.                 **if** $p \in \mathcal{T}_i$ **then**
13.                     $C_i[p] = C_i[p] - (C_l[e] - \alpha_2)$ (conservatively remove residual)
14.                 **end if**
15.             **end for**
16.         **end if**
17.     **end for**
18. **end for**
19. Output $\left(\mathcal{S}, \tilde{f}_X(\mathcal{S})\right)$.

Algorithm 4: Final Algorithm

**Corollary 4.1.2.1 (Privacy of Final Algorithm)** **Algorithm 4** is $(\varepsilon, \delta)$-DP

The proof follws from that the fact the output is post-processing the $(\varepsilon, \delta)$-DP output of **Algorithm 3**.

**Theorem 4.1.3 (Error)** For all $p \in \mathcal{S}$, we have with probability $1 - \beta$,

$$|\tilde{f}_X(p) - f_X(p)| \leq \alpha$$

where

$$\alpha = \left(1 + \frac{6h \log(3h/\delta)}{\varepsilon}\right) + \frac{n}{\kappa + 1} + \left(\frac{8h}{\varepsilon} \log\left(\frac{2\kappa h}{\beta}\right)\right)$$

*Proof.* Next we state the well known estimation error due to space constraints for the Misra Gries algorithm. A proof for this lemma can be found in [4:Section 2].

**Lemma 4.1.4 (MG Error)** Let $C[x]$ denote the frequency estimate for $x \in X$, given by a MG sketch with $\kappa$ counters on input stream $X$. Then $C[x] \in \left[f_X(x) - \frac{n}{\kappa+1}, f_X(x)\right]$.

For any fixed level of the hierarchy, with probability $1-\beta/h$ we have three sources of error. The first source of $1+\frac{6h\log(3h/\delta)}{\varepsilon}$ comes from suppressing isolated counts. The second error term $\frac{n}{\kappa+1}$ comes from the error of the deterministic MG counter (**Lemma 4.1.4**). For the third error term, for some $b$ to be defined later, at each level $l \in [h]$, we want with probability at most $\beta/2h$ that $|\gamma_l| > b/2$ and with probability at most $\beta/2h$ we want for $i \in [\kappa]$, $|w_i| > b/2$. This guarantees, with probability $1-\beta/h$, that $\max_{i\in\kappa}|w_i| + |\gamma_l| \leq b$. Using the tail bounds of the Laplace distribution, and taking union bounds over all the counters we get

$$\Pr\left[\max_{i\in\kappa} w_i > b/2\right] \leq \kappa e^{-\frac{\varepsilon}{8h}b} \tag{50}$$

Setting $\beta/2h = \kappa e^{-\frac{\varepsilon}{8h}b}$, we get

$$b \geq \frac{8h}{\varepsilon}\log\frac{2\kappa h}{\beta} \tag{51}$$

We can also bound the error from $\gamma_l$ using a similar analysis, and it turns out the above value for $b$ suffices to bound the error due to $\gamma_l$. Finally, we take the union bound over all $h$ levels, and get with probability $1-\beta$, the total error incurred by *any* node is at most

$$\alpha = \left(1+\frac{6h\log(3h/\delta)}{\varepsilon}\right) + \frac{n}{\kappa+1} + \left(\frac{8h}{\varepsilon}\log\left(\frac{2\kappa h}{\beta}\right)\right) \tag{52}$$

□

We note that the dependence on $h$ here is not optimal. We have used basic composition to show a linear dependence on $h$, in order to keep the development clear. However, it is possible to show an improved dependence on $\sqrt{h}$, by invoking more advanced composition theorems [5, 20]. Since we assume that $h$ is relatively small in practice, we don't expand on this point in this presentation. Note that as we cannot avoid the composition error due to the height of the hierarchy, we can directly estimate the unconditional frequencies of a node $p$ by looking at the noisy count of $C_{\mathsf{Level}(p)}[p]$. However, this means that there is no advantage to reporting relative error guarantees instead of the absolute error in the streaming case: all prefixes incur noise of the same magnitude. We leave it open to show whether this gap is provably unavoidable or can be surmounted using different techniques.

**Theorem 4.1.5 (Coverage)** For all $p \notin \mathcal{S}$, with probability $1-\beta$, $F_{X,\mathcal{S}}(p) \leq \tau - \alpha$.

*Proof.* Let $\mathcal{S}$ denote the output of **Algorithm 4**. Fix $p \notin \mathcal{S}$, define $\mathcal{A}_p = \{q \in \mathcal{S} : q \succ p \wedge \nexists q' \text{ s.t. } q \succ q' \wedge q' \succ p\}$. Let $s_p$ denote the amount of weight removed from the unconditional count of $p$ in the removal step of **Algorithm 4** for prefix $p$. Note that $s_p \leq \sum_{q\in\mathcal{A}_p} f_X(q)$ by definition, as MG counts cannot ever overestimate $f_X(e)$. We need to account any overestimation due to differential privacy. With probability $1-\beta$, we have that the noise per node due to DP is $\alpha_2$. Thus, with probability $1-\beta$, $C_l[e]-\alpha_2$ in the removal line is strictly less than $f_X(e)$ (as we have removed the DP contribution and $C_l[e] \leq f_X(e)$). So we always take off less than we could from the count of $p$. Any prefix $p$ is included in $\mathcal{S}$ only if $C_l[p] + \alpha_1 > \tau - \alpha_1$. So for all $p \notin \mathcal{S}$, we must have $C_l[p] + \alpha_1 - s_p \leq \tau - \alpha_1$. Therefore, with probability $1-\beta$,

$$F_{X,\mathcal{S}}(p) = f_X(p) - \sum_{q\in\mathcal{A}_p} f_X(q) \tag{53.1}$$

$$\leq (C_l[p] + \alpha_1) - s_p \tag{53.2}$$

$$\leq \tau - \alpha_1 \tag{53.3}$$

**Equation 53.1** comes from the definition of residual count. **Equation 53.2** comes from the fact that $s_p \leq \sum_{q \in \mathcal{A}_p} f_X(q)$, and $f_X(p) \leq C_l[p] + \alpha_1$ with probability $1 - \beta$ (from **Theorem 4.1.3**). □

## A Composition and Structure

Differential Privacy was initially motivated by the study of counting queries, and heavy hitter estimation can be seen as post processing of private release of counting queries. The privacy community has studied extensively the composition of privacy parameters when dealing with *arbitrary* counting queries. From basic composition we know that the maximum DP error of $q$ invocations of the Laplace mechanism scales $O(1/\varepsilon \cdot q \log[q])$. Thomas Steinke and Jonathan Ullman. [40] show that we can save the $\log q$ factor by exploiting the structure of correlated noise used for DP, and have error scale by $O(q/\varepsilon)$. Of course we could also use advanced composition [21] on top of this to further reduce the error to $O\Big(1/\varepsilon \cdot \sqrt{q \log[1/\delta]}\Big)$. Peter Kairouz, Sewoong Oh, and Pramod Viswanath. [27] show an exact characterisation of the best privacy parameters that can be guaranteed when composing many $(\varepsilon, \delta)$-differentially private mechanisms. Unfortunately computing these parameters for arbitrary queries with different privacy parameters is #P-complete [37]. While the early work was focused on arbitrary counting queries, there has been considerable work in exploiting structure in queries to obtain better error than basic or advanced composition. Cynthia Dwork, Moni Naor, Omer Reingold, Guy N Rothblum, and Salil Vadhan. [19] proposed the Above Threshold algorithm and showed that one can group similar queries together, and replace them with a single query to pay for many queries just once. This insight has led to multiple constructions of highly accurate DP algorithms that would have been impractical if we considered basic or advanced composition [7, 9, 17, 38]. Haim Kaplan, Yishay Mansour, and Uri Stemmer. [28] identified that for certain distribution of queries, the composition of the above threshold algorithm could be further improved, by observing that not all large queries include the neighbouring element. Wei Dong, Dajun Sun, and Ke Yi. [16] show how to bypass composition for special class of conjunctive queries. We direct the reader to the chapter by Thomas Steinke. [41] for a detailed survey on the role of composition in differential privacy. Despite an enormous body of work on counting queries and composition, the question of hierarchical counting with privacy has remained unexplored. In this work, we show that one can use similar tricks to the above work to exploit the structure of a hierarchy, and get highly accurate algorithms, that are not possible with general purpose techniques.

## B Greater Privacy For Larger Groups

As discussed earlier, we only bound the relative error of approximating the unconditional frequency of any prefix. As pointed out by [23:Page 2], if we instead wanted to upper bound the worst case absolute error, then it is known that the error *must* scale linearly with the height of the hierarchy. The problem of estimating private counts in a tree can be reformulated as releasing a *linear query* of the form $A\vec{x}$ privately, where $\vec{x}$ is a vector of unconditional counts for all leaves in the hierarchy, and then finding heavy hitters post processing. The matrix $A \in \{0,1\}^{|\mathcal{H}| \times |\mathcal{H}|}$ is an adjacency matrix representation of the tree that represents if one node is a parent of another or not. Releasing linear queries privately has been exhaustively studied in the privacy community [18, 22, 25, 44]. Alexander Edmonds, Aleksandar Nikolov, and Jonathan Ullman. [22] provide lower bounds for the absolute error of linear queries under pure and approximate differential privacy. They show that even when considering $(\varepsilon, \delta)$-DP, the absolute error for any private linear query algorithm scales $\|A\|_\infty = \Theta(h)$ (which is the same as just releasing the entire tree privately with Laplace noise). Remember, we are able to circumvent composition by leveraging the structure of a hierarchy. Requiring the error of every estimate at every level to be bounded by the same constant destroys this structure. We cannot use information gained lower down the hierarchy to make useful claims about elements higher in the hierarchy. We need to treat each level independently as we want the noise for each node to be independent. Thus the advantage of our algorithm, and that of [23:Algorithm 1], is that we can re-use information from earlier queries while still preserving privacy, paying for absolute error instead. A second advantage to considering relative error is that it is the more practical notion of error. This was also observed by David Pujol and Damien Desfontaines. [39], who posted as an open problem the task of finding algorithms that allow larger groups to have more privacy than smaller groups. When dealing with hierarchies, nodes higher up in the tree by definition can be estimated by counts lower in the tree (via partitions or hierarchical heavy hitters) — and this information is public knowledge. So we would like to utitlise this when designing algorithms. Consider the situation where the exact count of a node is $10^6$. If we incur an absolute error of 100 units when estimating the count of a node, we do not expect that to affect the final social decision associated with this private statistic. However, when the exact answer is 99, say, and we incur an error of 100 units, such a large error in estimation is unacceptable.

## C Relationship To Counting Over Trees

The estimation error from **Theorem 3.6.2** can be used to obtain better results in practice for [23:Definition 1.4] for the simpler problem of estimating unconditional frequencies with constant relative error. Note the main difference between our result and that of Badih Ghazi, Pritish Kamath, Ravi Kumar, Pasin Manurangsi, and Kewen Wu. [23] is that they add noise with scale $O(c/\varepsilon)$ to estimate the count of any node, regardless of the distribution of heavy hitters[12]. In our case, the relative error is independent of both the height and the number of hierarchical heavy hitters. Furthermore, our algorithm is simpler to define. In [23:Algorithm 3], the authors use a geometric progression of decreasing $\tau$'s, and repeatedly apply [23:Algorithm 1] on these thresholds to get the relative error guarantee. It is not clear how to set these thresholds for practical algorithms on real-world datasets, and the constants are larger than the height of any hierarchical domain in practice.

## D A Note About Re-using Random Samples

In this section we describe how the privacy proof analysis breaks down if we re-use the same random samples for thresholding, and outputting the counts. Borrowing notation from the proof of **Theorem 4.1.2**, if we used fresh randomness then we can go from

$$\Pr_{\vec{w},\gamma}[C + \gamma 1^{\kappa} + \vec{w} = \vec{y}] = \Pr_{\gamma}[C + \gamma 1^{\kappa} + \vec{w} = \vec{y} \mid \vec{w}] \Pr[\vec{w}] \tag{54}$$

to

$$= \Pr[\vec{w}] \prod_{i=1}^{\kappa} \Pr_{\gamma}[C[i] + \gamma + w_i = y_i \mid w_i] \tag{55}$$

without any issues. The noisy released count of $C[i]$ is independent of the thresholding operation, and the proof holds.

On the other hand, if we reused noise then the second equation has further constraints. Consider the event that all noisy counters are above the threshold (this is the worst case, where we reveal maximal information about $\gamma$), and let $X_i = C[i] + \gamma + w_i$ for $i \in [\kappa]$. Then assuming we release counts in lexicographical order, we have

$$\Pr_{\vec{w},\gamma}[C + \gamma 1^{\kappa} + \vec{w} = \vec{y}] = \Pr[\vec{w}] \prod_{i=1}^{\kappa} \Pr_{\gamma}\Big[X_i = y_i \mid \vec{w} \wedge_{j<i} X_j = \widetilde{C[j]} \wedge_{j<i} X_j \geq \tau\Big]. \tag{56}$$

[12] Remember $c$ is an upper bound on the total number of hierarchical heavy hitters.

Each release of a noisy count puts a constraint on the possible values of $\gamma$. Thus, we are no longer able to just use the pdf of the Laplace distribution to upper bound the ratios as we did in our proof (because the privacy distribution has changed to a truncated version of Laplace distribution). Viewing the same algorithm with the SVT lens, [34:Page 5, Algorithm 3] argue how not using a fresh batch of randomness is *not* differentially private by showing how a constraint on the value of $\gamma$ needs to be ignored to complete the privacy proof (equation 11 on Page 5). This issue was originally pointed out by [43:Appendix A] where they show that re-using randomness puts additional constraints on the support of the randomness, which results in the variance of the privacy distribution to shrink. To make up for this shrinkage, they show that the scale of the noise distribution would need to be linear with the number of queries. This destroys any benefit to using SVT in the first place.